\documentclass{llncs}
\usepackage[usenames]{color}
\usepackage[T1]{fontenc}
\usepackage{amsmath}
\usepackage{amssymb}
\usepackage{paralist}
\usepackage{wrapfig}
\usepackage{stackrel}
\usepackage{adjustbox}
\usepackage{booktabs}
\usepackage{subcaption}
\usepackage[linesnumbered,ruled,noend]{algorithm2e}

\usepackage[only,nnearrow,llbracket,rrbracket,Lbag,Rbag,llparenthesis,rrparenthesis]{stmaryrd}
\usepackage{trimclip}

\usepackage[english]{babel}
\usepackage{blindtext}

\usepackage{enumitem}
\usepackage{braket}
\usepackage{comment}
\usepackage{siunitx}
\usepackage{scalerel}
\SetKwProg{Fun}{}{ is}{end}
\usepackage{tikz}
\usetikzlibrary{automata,arrows,shapes.misc,backgrounds,positioning,calc}
\usepackage{tikz-qtree}
\usepackage{pgfplots}
\pgfplotsset{compat=1.3}
\usepgfplotslibrary{groupplots}
\tikzset{cross/.style={cross out, draw=black, minimum size=2*(#1-\pgflinewidth), inner sep=0pt, outer sep=0pt}, cross/.default={1pt}}

\InputIfFileExists{synctex.tex}{%
\synctex=1
}{%
}

\InputIfFileExists{cutmargins.tex}{%
	\pdfpagesattr{/CropBox [92 112 523 758]} % LNCS page made big 
}{%
}

\definecolor{darkpastelgreen}{rgb}{0.01, 0.75, 0.24}

\makeatletter
\DeclareRobustCommand{\shortto}{%
  \mathrel{\mathpalette\short@to\relax}%
}

\DeclareRobustCommand{\shortminus}{%
  \mathrel{\mathpalette\short@minus\relax}%
}

\newcommand{\short@to}[2]{%
  \mkern2mu
  \clipbox{{.5\width} 0 0 0}{$\m@th#1\vphantom{+}{\rightarrow}$}%
  }

\newcommand{\short@minus}[2]{%
  \mkern2mu
  \clipbox{{.5\width} 0 0 0}{$\m@th#1\vphantom{+}{-}$}%
  }
\makeatother

\newcommand{\labeledto}[1]{{{\shortminus}\hspace{-2pt}\raisebox{0.16ex}{$\scriptstyle\{ #1\hspace{-0.28pt}\}$}\hspace{-2.2pt}{\shortto}}}
\newcommand{\scriptlabeledto}[1]{{{\shortminus}\hspace{-1.0pt}\raisebox{0.12ex}{$\scriptscriptstyle\{ #1\hspace{-0.28pt}\}$}\hspace{-1.6pt}{\shortto}}}
\def\move(#1,#2,#3){%
\mathchoice
{#1\,\labeledto{#2}\,#3}
{#1\labeledto{#2}#3}
{#1\scriptlabeledto{#2}#3}
{#1\scriptlabeledto{#2}#3}
}

\def\markedmove(#1,#2,#3,#4){#1 \stackrel{#2}{\rightarrow}_{#4} #3}
\def\minterms#1{\mathit{Minterms}(#1)}
\newcommand{\minterm}{\omega}
\newcommand{\issatfn}{\mathit{IsSat}}
\def\issat#1{\issatfn(#1)}
\newcommand{\card}[1]{|{#1}|}

\newcommand{\bigO}[1]{\mathcal{O}\big(#1\big)}

\newcommand{\Csharp}{\settoheight{\dimen0}{C}C\kern-.05em \resizebox{!}{\dimen0}{\raisebox{\depth}{\#}}}
\newcommand{\Cpp}{C\nolinebreak\hspace{-.03em}\raisebox{.3ex}{+}\nolinebreak\hspace{-.07em}\raisebox{.3ex}{+}}

\newcommand{\algebra}{\mathcal{A}}
\newcommand{\domain}{\mathfrak{D}}

\newcommand{\simul}{\preceq}

\newcommand{\Rel}{\mathit{Sim}}
\newcommand{\NotRel}{\mathit{NotSim}}
\newcommand{\Rm}{\mathit{Rm}}

\newcommand{\BDD}[1]{\mathbf{BDD}_{#1}}

\newcommand{\lang}{\mathcal{L}}

\newcommand{\goabove}[3]{#1{\raisebox{+0.7ex}{\scaleobj{0.6}{#2}}\!\!\!\nnearrow}#3}
\newcommand{\denote}[1]{\llbracket{#1}\rrbracket}

\newcommand{\ML}{M_{L}}

\newcommand\graphwidth{4.8cm}
\newcommand\graphwidthother{6cm}

\makeatletter
\let\original@algocf@latexcaption\algocf@latexcaption
\long\def\algocf@latexcaption#1[#2]{%
  \@ifundefined{NR@gettitle}{%
    \def\@currentlabelname{#2}%
  }{%
    \NR@gettitle{#2}%
  }%
  \original@algocf@latexcaption{#1}[{#2}]%
}
\makeatother

\definecolor{maroon}{rgb}{0.5, 0.0, 0.0}

\newcommand{\ol}[1]{\textcolor{blue}{\ifmmode \text{[OL: #1]}\else [OL: #1] \fi}}
\newcommand{\lh}[1]{\textcolor{blue}{\ifmmode \text{[LH: #1]}\else [LH: #1] \fi}}
\newcommand{\tv}[1]{\textcolor{magenta}{\ifmmode \text{[TV: #1]}\else [TV: #1] \fi}}
\newcommand{\js}[1]{\textcolor{green}{\ifmmode \text{[JS: #1]}\else [JS: #1] \fi}}
\newcommand{\margus}[1]{\textcolor{maroon}{\ifmmode \text{[MV: #1]}\else [MV: #1] \fi}}

\renewcommand{\lh}[1]{}
\renewcommand{\tv}[1]{}
\renewcommand{\js}[1]{}

\newcommand{\vata}[0]{\textsc{Vata}\xspace}
\newcommand{\mona}[0]{\textsc{Mona}\xspace}
\newcommand{\dwina}[0]{\textsc{dWiNA}\xspace}
\newcommand{\regex}[0]{\textsc{RegEx}\xspace}
\newcommand{\wsones}[0]{\textsc{WS1S}\xspace}
\newcommand{\asgn}[0]{:=}
\newcommand{\post}[0]{\Delta}
\newcommand{\postofin}[3]{{#3}(#2,#1)}
\newcommand{\postof}[2]{\post(#2,#1)}

\newcommand{\transml}[0]{\Delta_{L}}
\newcommand{\algoname}[1]{\text{\textsc{#1}}\xspace}
\newcommand{\inysim}[0]{\algoname{INY}}
\newcommand{\globalsim}[0]{\algoname{GlobINY}}
\newcommand{\nocountsim}[0]{\algoname{NoCount}}
\newcommand{\localsim}[0]{\algoname{LocalMin}}
\newcommand{\gmrtsim}[0]{\algoname{GlobRT}}
\newcommand{\rtsim}[0]{\algoname{RT}}
\newcommand{\hhksim}[0]{\algoname{HHK}}

\newcommand{\sct}[1]{\S{\ref{#1}}}
\newcommand{\run}{\rho}
\renewcommand{\phi}{\varphi}
\newcommand{\Predicates}{\mathbb{P}}
\newcommand{\predset}{\Phi}
\newcommand\alphabetof[1]{\Predicates_{#1}}

\newcommand{\moves}[2]{{#1}\rightarrow{#2}}
\newcommand{\Reach}[2]{\Gamma(#1,#2)}

\newcommand{\eqdef}{\stackrel{\raisebox{-1mm}{\textrm{\tiny def}}}{=}}

\newcommand{\compl}[1]{#1^{\complement}}

\newcommand{\Csat}{\mathcal{C}_{\textit{sat}}}

\newcommand{\maxoutdeg}[0]{W}

\newcommand{\notsim}[1]{\npreceq_{#1}}
\title{
  Simulation Algorithms for Symbolic Automata
  (Technical Report)
}

\author{
  Luk\'{a}\v{s} Hol\'{i}k\inst{1},
  Ond\v{r}ej Leng\'{a}l\inst{1}, 
  Juraj S{\'i}{\v c}\inst{1,2},\\
  Margus Veanes\inst{3}, and
  Tom\'{a}\v{s} Vojnar\inst{1}
}

\institute{
  {FIT, Brno University of Technology, IT4Innovations Centre of Excellence,
   Czech~Republic}
  \and
  {Faculty of Informatics, Masaryk University, Brno, Czech~Republic}
  \and
  {Microsoft Research, Redmond, USA}
}

\begin{document} 
%%%%%%%%%%%%%%%%%%%%%%%%%%%%%%%%%%%%%%%%%%%%%%%%%%%%%%%%%%%%%%%%%%%%%%%%%%%%%%%%

%HERE SPACES SMALLER SPACES AROUND ALGORITHMS AND TABLES!!!!!!!!!!
\setlength{\textfloatsep}{10pt plus 1.0pt minus 2.0pt}
\setlength{\floatsep}{10pt plus 1.0pt minus 2.0pt}
\setlength{\intextsep}{10pt plus 1.0pt minus 2.0pt}

\maketitle

%\tv{The title ``Simple Algorithms...'' is not much enthusiastic. Sounds like we
%are ourselves afraid that the contribution is small. I better switched to just
%``Efficient Computation of ...''. If you don't like it, we can think of
%something like ``Simple and/but Efficient...'' or perhaps just ``Computing
%Simulation...''?}
%\lh{don't like Efficient. "Simple" might be a good thing :). They are simple in comparison with RT and like. 
%  Or just the previous "Computing ..."?}
%\ol{I'm not so wild about ``simple,'' too.
%How about putting the main observation into the title. I.e. sth like (stupid suggestions): ``Keep it symbolic: on computing simulation over symbolic automata.''
%or ``No Counters for Old Men: On Computing Simulation over Symbolic Automata''}

%\vspace{-4mm}
\begin{abstract}

  We investigate means of efficient computation of the simulation relation over
  symbolic finite automata (SFAs), i.e., finite automata with transitions
  labeled by predicates over alphabet symbols.
  In one approach, we build on the algorithm by Ilie, Navaro, and Yu proposed
  originally for classical finite automata, modifying it using the so-called
  mintermisation of the transition predicates.
  This solution, however, generates all Boolean combinations of the
  predicates, which easily causes an exponential blowup in the number of transitions.
  Therefore, we propose two more advanced solutions.
  The first one still applies mintermisation but in a local way, mitigating the
  size of the exponential blowup.
  %
  % The second one focuses on a~more principled approach of combining predicates,
  The other one focuses on a~novel symbolic way of dealing with transitions,
  for which we need to sacrifice the counting technique of the original
  algorithm (counting is used to decrease the dependency of the running time on the number
  of transitions from quadratic to linear).
  We perform a thorough experimental evaluation of all the algorithms, together
  with several further alternatives, showing that all of them have their merits
  in practice, but with the clear indication that in most of the cases,
  efficient treatment of symbolic transitions is more beneficial than
  counting.

  % We experimentally justify the practical usefulness of simulations for
  % symbolic automata and efficiency of our algorithms in comparison with
  % standard automata algorithms when run after mintermisation.
%
%  We also observe that focusing on optimising treatment of symbolic
%  transitions, as done in our second algorithm, appears overall the most
%  efficient.
\end{abstract}

%!!!!!!!!!!!!!!!!!!!!!!!!!!!!!!!!
%\enlargethispage{3mm}
%!!!!!!!!!!!!!!!!!!!!!!!!!!!!!!!!

%%%%%%%%%%%%%%%%%%%%%%%%%%%%%%%%%%%%%%%%%%%%%%%%%%%%%%%%%%%%%%%%%%%%%%%%%%%%%%%%
%\vspace{-10.0mm}
\section{Introduction}\label{sec:label}
%\vspace{-2.0mm}
%%%%%%%%%%%%%%%%%%%%%%%%%%%%%%%%%%%%%%%%%%%%%%%%%%%%%%%%%%%%%%%%%%%%%%%%%%%%%%%%

We investigate algorithms for computing simulation relations on states of
symbolic finite automata.
\emph{Symbolic finite automata} (SFAs)~\cite{W99,Veanes:2012:SAT:2260516.2260560} extend the
classical (nondeterministic) finite automata (NFAs) by allowing one to annotate a transition with
a~predicate over a~possibly infinite alphabet.  
Such \emph{symbolic} transitions then represent a set of all (possibly
infinitely many) concrete transitions over all the individual symbols that
satisfy the predicate.
SFAs offer a practical solution for automata-based techniques whenever the
alphabet is prohibitively large to be processed with a standard NFA, for
instance, when processing Unicode-encoded text (e.g., within various
security-related analyses) or in automata-based decision procedures for logics
such as MSO or WS1S~\cite{DBLP:conf/wia/Veanes13,DV17}.
Applications of SFAs over arithmetic alphabets and formulas
arise also when dealing with symbolic transducers in the context of
various sanitizer and encoder analyses~\cite{DV17}.

A~\emph{simulation relation} on an automaton underapproximates the inclusion of
%languages of individual~states~\cite{DBLP:conf/cade/BustanG00,todo,todo,todo}.
languages of individual~states~\cite{Bustan:2003:SM:635499.635502}.
This makes it useful for reducing nondeterministic automata and in testing
inclusion and
equivalence of their languages~\cite{Bustan:2003:SM:635499.635502,DBLP:conf/tacas/AbdullaCHMV10,Bonchi:2013:CNE:2429069.2429124}.
Using simulation for these purposes is often the best compromise between two
other alternatives: \begin{inparaenum}[(i)] \item  the cheap but strict
bisimulation and \item  the liberal but expensive language inclusion.
\end{inparaenum}

The obvious solution to the problem of computing simulation over an SFA is to
use the technique of \emph{mintermisation}:
the input SFA is transformed into a form in which 
predicates on transitions partition the alphabet.
Predicates on transitions can then be treated as ordinary alphabet symbols and
most of the existing algorithms for NFAs can be used out of the box, including a
number of algorithms for computing simulations.
We, in particular, consider mintermisation mainly together with the algorithm by
Ilie, Navaro, and Yu from~\cite{INY} (called \inysim in the following), and, in
the experiments, also with the algorithm by Ranzato and Tapparo (called also \rtsim)~\cite{RT}.
A fundamental problem is that
% , in the theoretical worst case, 
mintermisation can increase the number of transitions exponentially due to
generating all Boolean combinations of the original transition predicates.
Moreover, this problem is not only theoretical, but causes a
significant blowup in practice too, as witnessed in the experiments
presented in this paper.

%\tv{I have put back that the above is not only a theoretical issue. I do think
%that it should be stressed for two reasons: (1) the blowup caused by the
%construction is really common and (2) we should prepare the grounds for the
%local mintermisation, which can also cause an exponential blowup, but not so
%easily/often.}
%
%% in the theoretical worst case, and is indeed expensive in practice, sometimes
%% prohibitively. 
%
%\tv{Note that I switched the order of m/n below. This is because it is much
%better to start with n in some of the more complex algorithms in order to get a
%readable bound. Hence, it is perhaps good to use the same order everywhere.}

%In order to cope with the above, we take as our starting point the \inysim
%algorithm, which has the best available time complexity $\bigO{nm}$ in terms of
%the number of states~$n$ and transitions~$m$ of the input~FA.
%
We therefore design algorithms that do not need mintermisation.
We take as our starting point the 
algorithm \inysim, which has the best available time complexity $\bigO{nm}$ in terms of
the number of states~$n$ and transitions~$m$ of the input~NFA.
We propose two generalisations of this algorithm.
The first one (called \localsim) reflects closely the ideas that \inysim uses to
achieve the low complexity.
Instead of applying \inysim on a globally mintermised SFA, it, however, requires
only a \emph{locally mintermised form}: for every state, the predicates on its
outgoing transitions partition the alphabet.
Local mintermisation is thus exponential only to the maximal out-degree of a
state.

Our second algorithm (called \nocountsim) is fundamentally different
from \localsim because it trades off the upfront mintermisation cost
against working with predicates in the algorithm, and therefore has a
different worst case computational complexity wrt the number of
transitions.  We show experimentally that this trade-off pays off.
To facilitate this trade-off, we had to drop a~counting technique that
 \inysim uses to improve its time complexity from $\bigO{n^2 m}$ to $\bigO{n m}$ and that
replaces repeated tests for existence of transitions with certain properties by
maintaining their number in a dedicated counter and testing it for zero.
Dropping the counter-based approach 
(which depends on at least local mintermisation) 
in turn allowed an additional optimisation
based on aggregating a~batch of certain expensive operations (satisfiability
checking) on symbolic transitions into one. 
Overall, this improves the
efficiency and ultimately reduces a worst-case $2^m$ cost, which is typically
\emph{independent} of the Boolean algebra, to the cost of inlining the Boolean
algebra operations, which may be polynomial or even (sub)linear in
$m$.

In our experiments, although each of the considered algorithms wins in some cases, our
new algorithms performed overall significantly better than \inysim with global
mintermisation.
\nocountsim performed the best overall, which suggests that avoiding
mintermisation and aggregating satisfiability tests over transition labels is practically
more advantageous than using the counting technique of~\inysim.
We have also compared our algorithms with a variant~\cite{holik:optimizing} of
the \rtsim algorithm, one of the fastest algorithms for computing simulation,
run on the globally mintermised automata (we denote the combination as
\gmrtsim).
The main improvement of \rtsim over \inysim is its use of partition-relation
pairs, which allows one to aggregate operations with blocks of the so-far
simulation indistinguishable states.
Despite this powerful optimisation and the fine-tuned implementation of \rtsim
in the \vata library \cite{LengalSV2012}, \nocountsim has a~better 
performance than \gmrtsim on automata with high diversity of transition predicates 
(where mintermisation significantly increases the number of transitions).
%

%\lukas{say something about that nondeterministic automata are useful?}

%\vspace{-3mm}
%--------------------------------------------------
\paragraph{Related work.}
Simulation algorithms for NFAs might be divided between \emph{simple} and
\emph{partition-based.}
Among the simple algorithms, the algorithm by Henzinger, Henzinger,
and K\"opke~\cite{HHK} (called \hhksim) is the first algorithm that achieved the time complexity
$\bigO{nm}$ on Kripke structures. 
The~later algorithm \inysim~\cite{INY} is a~small modification of \hhksim and
works on finite automata in a time at worst $\bigO{nm}$. The automata are
supposed to be complete (every state has an~outgoing transition for every
alphabet symbol).
INY can be adapted for non-complete automata by adding an
initialisation step which costs $\bigO{\ell n^2}$ time where $\ell$~is the size
of the alphabet, resulting in $\bigO{nm + \ell n^2}$ overall complexity
(cf.~\sct{sec:sim_nfa}).

The~first partition-based algorithm was \rtsim, proposed in~\cite{RT}.
The~main innovation of \rtsim is that the overapproximation of the simulation relation is
represented by a so-called \emph{partition-relation pair}.
In a~partition-relation pair, each class of the partition of the set of states
represents states that are simulation-equivalent in the current
approximation of the simulation, and the relation on the partition denotes the
simulation-bigger/smaller classes.
% A~partition-relation represents a pair is a relation on blocks of potentially
% simulation-equivalent states.
Working with states grouped into blocks is faster than working with individual
states, and in the case of the most recent partition-based algorithms for Kripke
structures~\cite{Cece17}, it allows to derive the time complexity $\bigO{n'm}$
 where $n'$~is the number of classes of the simulation
equivalence (the partition-based algorithms are also significantly faster in
practice, although their complexity in terms of~$m$ and~$n$ is
still~$\bigO{nm}$).
See e.g.~\cite{Cece17} for a~more complete overview of algorithms for computing simulation over NFAs
and Kripke structures.

Our choice of \inysim over \hhksim among the simple algorithms is justified by
a~smaller dependence of the data structures of \inysim on the alphabet size.
The main reason for basing our algorithms on one of the simple
algorithms is their relative simplicity.
Partition-based algorithms are intricate as well as the proofs of their small
asymptotic complexity.
Moreover, they compute predecessors of dynamically refined blocks of
states via individual alphabet symbols, which seems to be a~problematic step to
efficiently generalise for symbolic SFA transitions.
Having said that, it remains true that the technique of representing preorders
through partition-relation pairs is from the high-level perspective orthogonal
to the techniques we have developed to generalise \inysim. 
Combining both types of optimisations would be a~logical continuation of
this work.
It is, however, questionable if generalising already very complex
partition-based algorithms, such as~\cite{Cece17,RT}, is the best way to
approach computing simulations over~SFAs.
Most of the intricacy of the partition-based algorithms aims at combining the counting technique with the partition-relation pairs.
Our experimental results suggest, however, that rather than using the counting technique,
it is more important to optimise the treatment of symbolic transitions and to avoid mintermisation.

Our work complements other works on generalising classical automata algorithms to SFAs, 
mainly the deterministic minimisation \cite{DBLP:conf/popl/DAntoniV14} and computing of bisimulation~\cite{DAntoni2017}.

%%%%%%%%%%%%%%%%%%%%%%%%%%%%%%%%%%%%%%%%%%%%%%%%%%%%%%%%%%%%%%%%%%%%%%%%%%%%%%%%
%\vspace{-3.0mm}
\section{Preliminaries}\label{sec:prelims}
%\vspace{-2.0mm}
%%%%%%%%%%%%%%%%%%%%%%%%%%%%%%%%%%%%%%%%%%%%%%%%%%%%%%%%%%%%%%%%%%%%%%%%%%%%%%%%

%\subsection{Effective Boolean Algebras and Symbolic Finite Automata} \label{SFAs}
%An effective Boolean algebra is a tuple $\algebra = (\domain, \Psi, \denote{\_}, \bot, \top, \vee, \wedge, \lnot)$
%An effective Boolean algebra is a tuple $\algebra = (\domain, \Psi, \denote{\_})$
% where $\domain$ is a non-empty set called the \emph{domain}, $\Psi$ is a set of \emph{predicates} closed under ${\vee}, {\wedge}, {\lnot}$  and $\denote{ \_ }: \Psi \rightarrow 2^{\domain}$ is the \emph{denotation function} where $\denote{ \bot } = \emptyset, \denote{ \top } = \domain$ and for all $\varphi, \psi \in \Psi, \denote{ \varphi \vee \psi } = \denote{ \varphi } \cup \denote{ \psi }, \denote{ \varphi \wedge \psi } = \denote{ \varphi } \cap \denote{ \psi }$ and $\denote{ \lnot \varphi } = \domain \setminus \denote{ \varphi }$. For $\varphi \in \Psi$, we write $\issat{\varphi}$ when $\denote{ \varphi } \neq \emptyset$ and say that $\varphi$ is \emph{satisfiable}.
%

Throughout the paper, we use the following notation: If $R \subseteq A_1
\times \cdots \times A_n$ is an $n$-ary relation for $n \geq 2$, then
$R(x_1,\ldots,x_{n-1}) \eqdef \{y \in A_n \mid R(x_1,\ldots,x_{n-1},y)\}$ for
any $x_1 \in A_1, \ldots, x_{n-1} \in A_{n-1}$.
\mbox{Let $\compl{R}\eqdef (A_1\times \ldots \times A_n)\setminus R$.}

\paragraph{Effective Boolean algebra.}
An \emph{effective Boolean algebra} is defined as a~tuple $\algebra = (\domain,
\Predicates, \denote{\cdot}, \vee, \wedge, \lnot)$ where $\Predicates$ is a set of
\emph{predicates} closed under predicate transformers
${\vee,\wedge}:\Predicates\times\Predicates\rightarrow\Predicates$ and ${\lnot}:\Predicates\rightarrow\Predicates$.
A first order interpretation (denotation) $\denote{ \cdot }: \Predicates \rightarrow
2^{\domain}$ assigns to every predicate of $\Predicates$ a subset of the
\emph{domain}~$\domain$ such that, for all $\varphi, \psi \in \Predicates$, it holds
that $\denote{ \varphi \vee \psi } = \denote{ \varphi } \cup \denote{ \psi },
\denote{ \varphi \wedge \psi } = \denote{ \varphi } \cap \denote{ \psi }$, and
$\denote{ \lnot \varphi } = \domain \setminus \denote{ \varphi }$.
For $\varphi \in \Predicates$, we write $\issat{\varphi}$ when $\denote{ \varphi }
\neq \emptyset$ and say that $\varphi$ is \emph{satisfiable}. 
The predicate $\issatfn$ and the predicate transformers ${\land}$, ${\lor}$, and ${\neg}$ must be effective (computable).
We assume that $\Predicates$ contains predicates $\top$ and $\bot$ with $\denote{\top} = \domain$ and $\denote{\bot} = \emptyset$.
%
%\lh{really use the predicate transformers or keep just syntactic logical connectives?}
%
Let $\predset$ be a subset of $\Predicates$.
%
%We say that $\predset$ \emph{covers} a subset $X$ of $\domain$ if
%$\bigcup_{\varphi\in\predset} \denote{\varphi}  \supseteq X$. 
%
If the denotations of any two distinct predicates in $\predset$ are disjoint,
then $\predset$ is called a \emph{partition} (of the set
$\bigcup_{\phi\in\predset}\denote{\phi}$). 
%
%
%Let $\predset = \{\varphi_1, \varphi_2, \dots ,\varphi_n\} \subseteq \Psi$ a non-empty finite set of predicates that covers $\domain$. A satisfiable predicate $\bigwedge_{i = 1}^{n}\psi_i$ where $\psi_i$ is either $\varphi_i$ or $\neg\varphi_i$ is called \emph{minterm} generated from $\predset$ and
%the set of all minterms generated from $\predset$ is denoted by $\minterms{\predset}$. 
%
The set $\minterms{\predset}$ of \emph{minterms} of a~finite set $\predset$ of predicates
% arises by taking all satisfiable predicates from
is defined as the set of all satisfiable predicates of
$\{\bigwedge_{\phi\in \predset'} \phi \land \bigwedge_{\phi\in\predset\setminus \predset'}\neg \phi  \mid \predset'\subseteq \predset\}$.
%\ol{is defined as $\{ \psi = \bigwedge_{\phi\in \predset'} \phi \land \bigwedge_{\phi\in\predset\setminus \predset'}\neg \phi \mid \issat \psi, \predset'\subseteq \predset \}$}
%
%We also say that the minterm $\psi$ is created from the predicate $\varphi \in \Psi$ if in the conjunction defining the minterm $\psi$, $\varphi$ was used in its non-negated form. 
%
Notice that every predicate of $\predset$ is equivalent to a~disjunction of
\mbox{minterms in $\minterms \predset$.}
% , and if $\predset$ covers $\domain$, then $\minterms{\predset}$ is the
% coarsest partition of $\domain$ that refines $\predset$ (in the sense that
% every element of $\minterms{\predset}$ is entailed by precisely one element of
% $\predset$).

%We are now prepared to define symbolic finite automata. Informally, these automata have transitions labelled by predicates which denote subsets of elements from $\domain_{\algebra}$. $\domain_{\algebra}$ has then the same role as the alphabet for FA, but because one predicate can denote potentially infinite set of domain elements, symbolic finite automata can be used with a big or infinite alphabet while keeping a compact form. 
%
%\begin{definition}
%

%Below, we denote by $\Cbool(x,y)$ the maximum size of the predicate created
%using $x$ operations on predicates of size at most $y$.
%%
%Moreover, we denote by  $\Csat(x)$ the complexity of checking the satisfiability
%of a predicate of size $x$.

%\tv{Please check the below. (1) Instead of $\Csat$ and $\Cbool$ always used in
%pair, we use just one of them. (2) Given the remaining time and the fact that
%making some more precise analysis is costly, I would suggest to not distinguish
%the cost of conjunctions, disjunctions, negations, etc.}

Below, we assume that it is possible to measure the size of the predicates of
the effective Boolean algebra $\algebra$ that we work with.
We denote by $\Csat(x,y)$ the worst-case complexity of constructing a predicate 
obtained by applying $x$ operations of $\algebra$
on predicates of the size at most $y$ and checking its satisfiability.

%\tv{$\Csat$ is to be used in the appendix too!!!}

\paragraph{Symbolic finite automata.}
We define a~\emph{symbolic finite automaton} (SFA) as a~tuple $M = (Q,
\algebra, \Delta, I, F)$ where $Q$ is a finite set of \emph{states}, $\algebra
  = (\domain, \Predicates, \denote{\cdot}, \vee, \wedge, \lnot)$
is an effective Boolean algebra, $\Delta
\subseteq {Q \times \Predicates \times Q}$ is a~finite \emph{transition relation}, $I
\subseteq Q$ is a~set of \emph{initial states}, and $F \subseteq Q$ is a set of
\emph{final states}.
%
%ARROW NOTATION DENOTES EXISTENCE OF A TRANSITION, NOT JUST A DIFFERENT NOTATION FOR THE TRIPPPLE:
%We denote by $\markedmove(q, \psi, p, M)$ that $(q, \psi, p) \in \Delta$ and by
%$\markedmove(q, a, p, M)$ that $\markedmove(q, \phi, p, M)$ such that
%$\phi\in\Psi$.
%We may write only $\move(q, \psi, p)$ and $\move(q,a,p)$ if $M$ is clear from
%the context.
%
An element $(q, \psi, p)$ of~$\Delta$ is called a \emph{(symbolic) transition}
and denoted by $\move(q,\psi,p)$.
We write $\denote {\move(q,\psi,p)}$ to denote the set $\{\move(q,a,p)\mid
a\in\denote{\psi}\}$ of \emph{concrete transitions} represented by
$\move(q,\psi,p)$,
%, %where $\move(q,a,p)$ denotes simply the triple $(q,a,p) \in Q \times \domain
%\times Q$, 
and we let $\denote\Delta \eqdef \bigcup_{{\move(q,\psi,p)}\in\Delta} \denote{\move(q,\psi,p)}$.
%
%We define the set of \emph{successors} of a state $q \in Q$ via a
%predicate $\psi \in \Predicates$ as the set $\postofin \psi q \Delta \eqdef \{ p \in
%Q \mid {\move(q,\psi,p)}\in\Delta \}$.
%
For $q\in Q$ and $a\in\domain$ let
$\Delta(q,a) \eqdef \{p\in Q\mid \move(q,a,p)\}$.

In the following, it is assumed that predicates of all transitions of an SFA
are satisfiable unless stated otherwise.
%
%
%We assume the set $\preofin \psi q \Delta$ of \emph{predecessors} of a state~$q
%\in Q$ via a predicate $\psi \in \Predicates$ to be defined analogically.
%
%We may drop the superscript $\Delta$ if no confusion arises.
%\ol{is $\preof \psi q$ used anywhere?}
%\tv{I added a formal definition of$\move(q,a,p)$. However, one should perhaps
%also add some intuition: it is a concretization of the symbolic transition or
%something like that.}
%
A~sequence $\run = q_0 a_1 q_1 a_2 \cdots a_n q_n$
with $\move(q_{i-1}\!,a_i,q_i)\in\denote\Delta$ for every $1\leq i \leq n$
is a~\emph{run} of $M$ over the word $a_1 \cdots a_n$.
% such that,
% for each $i:1\leq i
% \leq n$, $\move(q_{i-1}\!,a_i,q_i)\in\denote\Delta$ is a \emph{run} of $M$ over the word $a_1
% \cdots a_n$.
The run~$\run$ is \emph{accepting} if $q_0 \in I$ and $q_n \in F$, and a word is
\emph{accepted} by $M$ if it has an accepting run.  The \emph{language}
$\lang(M)$ of $M$ is the set of all words accepted by~$M$.

An SFA $M$ is \emph{complete} iff, for all $q \in Q$ and $a \in \domain$, there
is $p\in Q$ with $\move(p,a,q)\in\denote\Delta$.
An SFA can be completed in a straightforward way:
from every state~$q$, we add a~transition from~$q$ labelled with
$\neg\bigvee\set{\phi \mid \exists p\in Q: 
\move(q, \varphi ,p) \in \Delta}$ to a~new non-accepting sink state, if the disjunction is satisfiable.
An SFA $M$ is \emph{globally mintermised} if the set
$\Predicates_\Delta \eqdef \{\phi\in\Predicates\mid \exists p,q:
\move(p,\phi,q)\in\Delta\}$ of the predicates appearing on its
transitions is a partition.
%, and it is \emph{local minterm normalised} if for every state $p\in Q$, the set $\Predicates_{\Delta,p} = \{\phi\in\Predicates\mid \exists q: \move(p,\phi,q)\in\Delta\}$ of the predicates appearing on the transitions starting from $p$ is a partition.
%
%
Every SFA can be made globally mintermised  
%and let $\minterms{\Psi_\Delta}$ be the set of all its \emph{minterms}, that is, the set $\{\varphi = \bigwedge_{\phi\in C} \phi \land \bigwedge_{\phi\in\Psi_\Delta\setminus C}\neg \phi \text{ s.t. } \issat{\varphi} \mid C\subseteq \Psi_\Delta\}$.
%
by replacing each $\move(p,\phi,q)\in \Delta$ by the set of transitions
$\{\move(p,\minterm,q)\mid \minterm \in \minterms{\Predicates_\Delta} \land
\issat{\minterm \land \phi}\}$ (see e.g. \cite{DBLP:conf/popl/DAntoniV14} for an efficient algorithm),
%\margus{Fixed: $\denote{\minterm} \subseteq \denote{\phi}$ is equivalent to
%               $\denote{\minterm}\cap \denote{\phi}\neq\emptyset$ because $\minterm$ is a minterm, thus 
%               $\issat{\minterm \wedge \phi}$ }
where $\issat{\minterm \wedge \phi}$ is an implementation of the test $\denote{\minterm} \subseteq \denote{\phi}$,
because if $\minterm$ is a minterm of $\Predicates_\Delta$ and $\phi\in\Predicates_\Delta$ then 
$\denote{\minterm}\cap \denote{\phi}\neq\emptyset$ implies that $\denote{\minterm} \subseteq \denote{\phi}$.
%
%Similarly, a local mintermised form is obtained by replacing every transition $\move(p,\phi,q)$ by the set of transitions 
%$\{\move(p,\minterm,q)\mid \minterm \in \minterms{\Predicates_{\Delta,p}} \land \denote{\minterm}\subseteq \denote{\phi}\}$.
%
Since for a~set of predicates $\predset$, the size of $\minterms \predset$ is at worst
$2^{|\predset|}$, global mintermisation is exponential in the number of
transitions. 
%Local mintermisation is considerably cheaper, it is only
%exponential to the largest number of transitions that start in a state.
%
%\lh{some key property of the global mintermized automaton that allows to use normal automata algorithms...?}

%%
%The set $\Predicates$ can then be seen as including $\Sigma$ and expressions
%built from it using $\vee$, $\wedge$, and $\neg$, which can be assigned the
%semantics of the set union, intersection, and complement (this is, however,
%purely formal as they are never used in $\Delta$).
%%

% !!!!!!!!!!!!!!!!!!!!!!!!!!!!!!!!!!!!!!!!!!!!!!!!!!!!!!!!!!!!!!!!!!!!!!!!!!!
%
% NOTE: cannot be easily moved before the above predicate since P_\Delta is
% defined only in the above paragraph.
%
% !!!!!!!!!!!!!!!!!!!!!!!!!!!!!!!!!!!!!!!!!!!!!!!!!!!!!!!!!!!!!!!!!!!!!!!!!!!

A~classical \emph{(nondeterministic) finite automaton} (NFA) $N = (Q, \Sigma,
\Delta, I, F)$ over a~finite alphabet $\Sigma$ can be seen as a special case of
an SFA where $\Delta$ contains solely transitions of the form $\move(q,a,r)$
s.t. $a\in\Sigma$ and $\denote a = \{ a \}$ for all $a\in\Sigma$.
Below, we will sometimes interpret an SFA $M = (Q, \algebra, \Delta, I, F)$
as its \emph{syntactic NFA} $N = (Q, \Predicates_\Delta, \Delta, I, F)$ in which 
the predicates \mbox{are treated as syntactic objects.}

%-----------------------------------------------------------
\paragraph{Simulation.} \label{sec:simulation}
Let $M = (Q, \algebra, \Delta, I, F)$ be an SFA. A relation $S$ on $Q$ is a
\emph{simulation} on~$M$ if whenever $(p,r) \in S$, then the following two conditions hold:
% \begin{inparaenum}%[label = (\roman*)]
\begin{inparaenum}[(C1)]
\item if $p \in F$, then $r \in F$, and \label{simFA1}
\item for all $a \in \domain$ and $p' \in Q$ such that $\move(p,a,p')\in\denote\Delta$, there is $r'\in Q$ such that $\move(r,a,r')\in\denote\Delta$ and $(p', r') \in S$. \label{simFA2}
\end{inparaenum}
There exists a unique maximal simulation on $M$, which is reflexive and
transitive. 
%\lh{\cite{} or say something?}. 
We call it the \emph{simulation (preorder)} on $M$ and denote it by ${\simul_M}$ (or $\simul$ when~$M$ is clear from the context).
%The subject of this paper is to design an algorithm that computes $\simul$ on a~given SFA.
Computing $\simul$ on a~given SFA is the subject of this paper.
A simulation that is symmetric is called a~\emph{bisimulation}, and \emph{the bisimulation equivalence} is the (unique) largest bisimulation, which is always an equivalence relation.
\section{Computing Simulation over SFAs}\label{sec:computing}
%\vspace{-2.0mm}
%%%%%%%%%%%%%%%%%%%%%%%%%%%%%%%%%%%%%%%%%%%%%%%%%%%%%%%%%%%%%%%%%%%%%%%%%%%%%%%%

In this section, we present our new algorithms for computing the simulation preorder over SFAs.
We start by recalling an algorithm for computing the
simulation preorder on an NFA of Ilie, Navarro, and Yu
from~\cite{INY} (called \inysim), which serves as the
basis for our work.
%
% Then, we introduce three algorithms for computing the simulation preorder over SFAs based on \inysim:
Then, we introduce three modifications of~\inysim for SFAs:
\begin{inparaenum}[(i)]
  \item  \globalsim,
  \item  \localsim, and
  \item  \nocountsim.
\end{inparaenum}
\globalsim is merely an application of the mintermisation technique: first globally mintermise
the SFA and then use \inysim to compute the NFA simulation preorder over the result.
%on its syntactic NFA.
%\ol{what's that?}
%
The main contribution of our paper lies in the other two algorithms,
which are subtler modifications of \inysim that avoid global mintermisation by reasoning 
about the semantics of transition predicates of~SFAs. 
%
%As shown by our experiments, each of them can be the fastest on some examples, but our algorithms, espesially \nocountsim, are the fastest most of the time.
%\ol{say none of them is strictly better than the other two}

%In this section, we start by recalling an algorithm for computing the
%simulation preorder on an NFA of Ilie, Navarro, and Yu
%from~\cite{INY} (called \inysim), which serves as the
%basis for our work.
%Then, we introduce three algorithms for computing the simulation over SFAs,
%which can all be seen as variants of \inysim:
%\begin{inparaenum}[(i)]
%  \item  \globalsim,
%  \item  \localsim, and
%  \item  \nocountsim.
%\end{inparaenum}
%\ol{say none of them is strictly better than the other two}

Before turning to the different algorithms, we start by explaining how
$\simul_M$ can be computed by an abstract fixpoint procedure and provide the
intuition behind how such a procedure can be lifted to the symbolic
setting.

\paragraph{Abstract procedure for computing~$\simul_M$.}
% We start by presenting an abstract fixpoint algorithm
% that we consider as the \emph{abstract procedure} for computing the
We start by presenting an \emph{abstract fixpoint procedure} for computing the
simulation~$\simul_M$ on an SFA $M = (Q, \algebra, \Delta, I, F)$.
We~formulate it using the notion of \emph{minimal nonsimulation} $\notsim{M}$
(which is a~dual concept to the maximal simulation~$\simul_M$ introduced
before), defined as the least subset~${\not \preceq} \subseteq Q \times Q$
s.t.\ for all $s,t \in Q$, it holds that
% First, the \emph{minimal nonsimulation} relation $\notsim{M}$ of $M$ is
% computed as the \emph{least
% subset ${\notsim{}}$} of $Q\times Q$ such that (\ref{eq:fixp}) holds:
%
\begin{equation}
  \label{eq:fixp}
  \begin{split}
    s \not\preceq t \mathrel{\Leftrightarrow} {} & (s \in F \land t
    \not\in F) \lor {} \\
    &  \exists i \in Q. \underbrace{\exists a \in \domain. (\move( s,a,i) \land \forall
      j \in Q. (\move(t,a,j)
      \Rightarrow i \not\preceq j))}_{(\textrm{\ref{eq:fixp}*})}.
  \end{split}
\end{equation}
%
% Then ${\notsim{M}}$ and ${\simul_M}\eqdef{\compl{\notsim{M}}}$ are well-defined
% because $Q$ is finite.
% It follows that $\simul_M$ is the unique \emph{maximal simulation} relation of
% $M$.
Informally, $s$ cannot be simulated by~$t$ iff
(line~1) $s$ is accepting and $t$ is not, or
(line~2) $s$ can continue over some symbol~$a$ into~$i$, while~$t$ cannot
simulate this move by any of its successors~$j$.
It is easy to see that ${\simul_M} = {\compl{\notsim{M}}}$.
The~algorithms for computing simulation over NFAs are efficient implementations
of such a fixpoint procedure using \emph{counter-based} implementations for
evaluating~(\ref{eq:fixp}*). 
Namely, for every symbol $a$ and a~pair of states~$t$ and~$i$, it keeps count
of those states~$j$ that could possibly contradict the universally quantified
property.
The~count dropping to zero means that the property holds universally.

\paragraph{Symbolic abstract procedure for computing~$\simul_M$.}
When the domain~$\domain$ is very large or infinite, then evaluating
(\ref{eq:fixp}*) directly is infeasible.
If
$\minterms{\Predicates_\Delta}$ is exponentially larger than the set
$\Predicates_\Delta$, then evaluating (\ref{eq:fixp}*) with $a$
ranging over $\minterms{\Predicates_\Delta}$ may also be infeasible.
Instead, we want to utilize the operations of the algebra~$\algebra$ without
explicit
reference to elements in $\domain$ and without constructing
$\minterms{\Predicates_\Delta}$.  The key insight 
%(used by algorithm \nocountsim) 
is that condition (\ref{eq:fixp}*)
is \mbox{equivalent to}% (\ref{eq:notsim-s})
\begin{equation}
  \label{eq:notsim-s}
\issat{\varphi_{si}\wedge\neg\Reach{t}{{\compl{\notsim{}}}(i)}}
\end{equation}
where, for $t,s,i\in Q$ and $J\subseteq Q$, we define
$\varphi_{si}\eqdef \bigvee_{(s,\psi,i)\in\Delta}\psi$ and 
$
\Reach{t}{J} \eqdef \bigvee_{j\in J} \varphi_{tj}
$,
i.e., $\Reach{t}{{\compl{\notsim{}}}(i)}$ is a~disjunction of predicates on all
transitions leaving~$t$ and entering a~state that simulates~$i$.
Using (\ref{eq:notsim-s})
to compute (\ref{eq:fixp}*) in the abstract procedure thus
eliminates the explicit quantification over $\domain$ and avoids
computation of $\minterms{\Predicates_\Delta}$.  The equivalence
between (\ref{eq:fixp}*) and (\ref{eq:notsim-s}) holds because, for
all $a\in\domain$ and $R\subseteq Q\times Q$, we have
%
%\begin{align*}
%  a\in\denote{\neg\Reach{t}{\compl{R}(i)}} 
%  &{}\Leftrightarrow
%  \neg\exists j(\move(t,a,j)\wedge (i,j)\in \compl{R}) \\
%  &{}\Leftrightarrow 
%  \forall j(\move(t,a,j) \Rightarrow (i,j)\in R).
%\end{align*}
%
\begin{equation*}
  a\in\denote{\neg\Reach{t}{\compl{R}(i)}} 
  \ \, \Leftrightarrow \ \,
  \neg\exists j(\move(t,a,j)\wedge (i,j)\in \compl{R}) 
  \ \, \Leftrightarrow \ \, 
  \forall j(\move(t,a,j) \Rightarrow (i,j)\in R).
\end{equation*}
The fixpoint computation based on (\ref{eq:notsim-s}) is used in our algorithm \nocountsim, which does not require mintermisation.
Its disadvantage is that it is not compatible with the counting technique.
Our algorithm \localsim is then a compromise between mintermisation and \nocountsim that retains the counting technique for the price of using a cheaper, local variant of mintermisation.

%*******************************************************************************
%\vspace{-3.0mm}
\subsection{Computing Simulation over NFAs (\inysim)}\label{sec:sim_nfa}
%\vspace{-2.0mm}
%*******************************************************************************

%\begin{wrapfigure}[15]{r}{72.5mm}
%\vspace*{-9mm}
%\hspace*{-3mm}
%\begin{minipage}{74.5mm}
  \begin{algorithm}[t]
    % \DontPrintSemicolon

    \KwIn{An NFA $N = (Q, \Sigma, \Delta, I, F)$}
    \KwOut{The simulation preorder $\simul_N$}

    % \lFor{$q \in Q, a \in \Sigma$} { \label{algFA:startInit}
    %     compute $\postof a q , \preof a q$
    % }

    \lFor{$p,q \in Q, a \in \Sigma$} { \label{algFA:startInit}
      $N_{a}(q,p) \asgn \card{\postof a q}$ \label{algFA:countinit}
    }

   % $\Rel \asgn {(Q \times Q) \setminus (F \times (Q \setminus F))}$\; \label{algFA:omega}
    %$\NotRel \asgn F \times (Q \setminus F)$\; \label{algFA:endInit}
    %$\Rel \asgn {Q \times Q}$\; \label{algFA:omega}
    %${\NotRel \asgn F \times (Q \setminus F)} \cup {\bigcup_{a\in\Sigma}\preof a Q \times (Q\setminus \preof a Q)}$\label{algFA:endInit}\

    $\Rel \asgn {Q \times Q}$\;
    %${\NotRel \asgn F \times (Q \setminus F)} \cup {\bigcup_{a\in\Sigma}\preof a Q \times (Q\setminus \preof a Q)}$\label{algFA:endInit}\
    ${\NotRel \asgn F \times (Q \setminus F)} \cup {\{(q,r)\mid \exists
    a\in\Sigma: \postof a q \neq \emptyset \land \postof a r = \emptyset\}}$\label{algFA:endInit}\;
    %$\Rel \asgn {(Q \times Q) \setminus \NotRel}$\; \label{algFA:omega}
    %$\Rel \asgn {(Q \times Q) \setminus \NotRel}$\; \label{algFA:omega}

    \While {$\NotRel \neq \emptyset$} {
      remove some $(i,j)$ from $\NotRel$ and $\Rel$\; \label{algFA:deq}
      \For{$\move(t,a,j) \in \Delta$} { \label{algFA:itersymbols}
        $N_{a}(t,i) \asgn N_{a}(t,i) - 1$\; \label{algFA:dec}
        \If(\tcp*[f]{$\goabove t {a} i = \emptyset$}){$N_{a}(t,i) = 0$} { \label{algFA:zero}
          \For {$\move(s,a,i) \in \Delta$ s.t. $(s,t) \in \Rel$} { \label{algFA:forLastStart}
           % $\Rel \asgn \Rel \setminus \{(s,t)\}$\;
            $\NotRel \asgn \NotRel \cup \{(s,t)\}$\; \label{algFA:enq}
          } \label{algFA:forLastEnd}
        }
      }
    }
    \Return $\Rel$\;
    \caption{\inysim}
    \label{alg:INY}
  \end{algorithm}
%\end{minipage}
%\end{wrapfigure}

%\ol{talk about the assumption that the NFA is complete}
%\lh{aaargh} 
In Algorithm~\ref{alg:INY}, we give a slightly modified version of the
algorithm $\inysim$ from~\cite{INY} for computing the simulation preorder over
an NFA $N = (Q,\Sigma,\Delta,I,F)$.
%
%We recall an algorithm for computing the simulation preorder $\simul$ over an NFA, or, in the other words,
%for computing the syntactic simulation preorder over an SFA. 
%$N = (Q, \algebra, \Delta, I, F)$, called
%\inysim~\cite{INY}, in Algorithm~\ref{alg:INY}.
%
The algorithm refines an overapproximation $\Rel$ of the simulation preorder until it
satisfies the definition of a~simulation.
%In the algorithm, $\Rel$ denotes the current approximation of~$\simul_N$.
%
The set $\NotRel$ is used to store pairs of states $(i, j)$ that were found to
contradict the definition of the simulation preorder. 
$\NotRel$~is initialised to contain (a)~pairs that contradict
condition~C\ref{simFA1}
%of a~simulation (cf.~\sct{sec:prelims})
and (b)~pairs that cannot satisfy condition~C\ref{simFA2} regardless of the
rest of the relation, as they relate states with incompatible outgoing symbols.
%, and then iteratively
%refines $\Rel$ until it is also consistent with condition~\ref{simFA2} of the
%definition.
%, it has been
%established that $i \not\simul j$ (initially filled with pairs of states not
%satisfying condition~\ref{simFA1}).
%
All pairs $(i,j)$ in $\NotRel$ are subsequently processed by removing
$(i,j)$ from $\Rel$ and propagating the change of $\Rel$ according
to~condition~C\ref{simFA2}:
for all transitions $\move(t,a,j)\in\Delta$,
it is checked whether $j$ was the last $a$-successor of~$t$ that could be
simulation-greater than~$i$
(hence there are no more such transitions after removing $(i,j)$ from $\Rel$).
If this is the case, then $t$ cannot simulate any $a$-predecessor $s$
of~$i$, and so all such pairs $(s,t)\in\Rel$ are added to $\NotRel$.
%Namely, if $t$ has no more $a$-moves above state~$i$, formally, $\goabove t {a}
%i \neq \emptyset$ where $\goabove t {a} i$ is the set $\{k \in Q \mid \move(t, a, k) \in
%\Delta \land (i, k) \in \Rel \cup \NotRel\}$.
%, the set of successors of~$t$ over~$a$
%that simulate~$i$ in the current approximation~$\Rel \cup \NotRel$).
%
%The algorithm keeps removing pairs $(i, j)$ from $\NotRel$ and processes
%each as follows: for all $t \in Q$ s.t. $\move(t,a,j)$ for some $a \in \alphabetof N$,
%check whether $t$ can go over symbol~$a$ above state~$i$, i.e., $\goabove t {a}
%i \neq \emptyset$
%(we use $\goabove t {a} i$ to denote the set $\{k \in Q \mid \move(t, a, k) \in
%\Delta \land (i, k) \in \Rel \cup \NotRel\}$, i.e., the set of successors of~$t$ over~$a$
%that simulate~$i$ in the current approximation~$\Rel \cup \NotRel$).
%\ol{current -> one of the previous?}
%If %$t$ cannot go over~$a$ above~$i$,
%not, then, for all $s$ s.t.\ $\move(s,a,i)$,
%it follows that $t$ cannot simulate $s$, i.e., $s \not\simul t$, so we remove
%$(s,t)$ from~$\Rel$ and add it into $\NotRel$.
%\ol{explain ``go above''}
%
%
%
%
In~order to have the previous test %($\goabove t {a} i \neq \emptyset$?)
efficient
(a~crucial step for the time complexity of the algorithm), the
algorithm uses a~three-dimensional array of counters~$N_a(t, i)$,
whose invariant at line~\ref{algFA:deq} is $N_a(t, i) = |\goabove t {a}
i|$ where  
$\goabove t {a} i$ is the set $\postofin a t \Delta \cap \Rel (i)$   
%$\{k \in Q \mid \move(t, a, k) \in \Delta \land (i, k) \in \Rel\}$ 
of successors of~$t$ over~$a$ that simulate~$i$ in the current
simulation approximation~$\Rel$.
In order to test $\goabove t {a} i = \emptyset$---i.e. the second conjunct of
(\ref{eq:fixp}*)---, it is enough to test if $N_a(t, i) = 0$.
% Thus $N_a(t, i) = 0$ means that
% $\postofin a t \Delta \cap \Rel (i) = \emptyset$
% which means that $\postofin a t \Delta \subseteq \compl{\Rel} (i)$ that corresponds
% to the second conjunct of (\ref{eq:fixp}*).

%Let $N = (Q, \Sigma, \Delta, I, F)$ be a~complete NFA with $n = |Q|$ states and
%$m = |\Delta|$ transitions. 
The lemma below shows the time complexity of
\inysim in terms of $n = |Q|$, $m = |\Delta|$, and $\ell = |\Sigma|$. 
The original paper~\cite{INY} proves the complexity $O(nm)$ for complete automata,
in which case $m\geq \ell n$, so the factor $\ell n^2$ is
subsumed by $nm$.
Since completion of NFAs can be expensive, the initialization step on
line~\ref{algFA:endInit} of our algorithm is modified (similarly as
in~\cite{nfasim}) to start with considering states with different sets of
symbols appearing on their outgoing transitions as simulation-different;
the cost of this step is
subsumed by the factor~$\ell n^2$ (see Appendix~\ref{app:inySimCompl} for the
proof of our formulation of the algorithm).
% in~\cite{nfasim}) to right away consider as simulation-different states with
% different sets of symbols appearing on their outgoing transitions with a cost
% subsumed by the factor $\ell n^2$ (see Appendix~\ref{app:inySimCompl} for the
% proof of our formulation of the algorithm).

%Since completion of NFAs is expensive, we add a preprocessing that allows to
%work with incomplete automata (similarly as in~\cite{nfasim}) with a cost
%subsumed by the factor $\ell n^2$ (see Appendix~\ref{app:inySimCompl} for the
%proof of our formulation of the algorithm).

%slightly different formulation of the algorithm, neglecting the factor $k$. A~proof for our formulation is provided in Appendix~\ref{app:inySimCompl}. 

\begin{lemma} \label{lemma:iny} \inysim computes~$\simul_N$ in time $\bigO{nm+\ell n^2}.$\end{lemma}

%We note that the $\bigO{n m}$ complexity crucially depends on our assumption that
%$N$~is complete (otherwise the size of the alphabet would have to be taken into
%account too).

%*******************************************************************************
%\vspace{-4.0mm}
\subsection{Global Mintermisation-based Algorithm for SFAs (\globalsim)}\label{sec:globalsim}
%\vspace{-0.0mm}
%*******************************************************************************

%\begin{wrapfigure}[6]{r}{62mm}
%\vspace*{-9mm}
%\hspace*{-3mm}
%\begin{minipage}{64mm}
  \begin{algorithm}[t]
    % \DontPrintSemicolon

    \KwIn{An SFA $M = (Q, \algebra, \Delta, I, F)$}
    \KwOut{The simulation preorder $\simul_M$}

    % compute global minterm normalized form of $M$, $\MG = (Q, \algebra, \Delta_{\MG}, I, F)$\;
    $\Delta_G \asgn$ globally mintermised $\Delta$\;

    \Return $\inysim((Q, \Predicates_{\Delta_G}, \Delta_G, I, F))$\;

    \caption{\globalsim}
    \label{alg:GlobalSim}
  \end{algorithm}
%\end{minipage}
%\end{wrapfigure}

The algorithm \globalsim (Algorithm~\ref{alg:GlobalSim}) is the initial
solution for the problem of computing the simulation preorder over SFAs.
It first globally mintermises the input automaton $M = (Q,
\algebra, \Delta, I, F)$, then
interprets the result as an NFA over the alphabet of the minterms,
and runs \inysim on the NFA. 
%effectively transforming it into an NFA~$N_M$ where every predicate $\psi$ occurring on
%a~transition $\move(p, \psi, q)$, for some states $p,q \in Q$, is treated as
%an uninterpreted symbol.
%After the transformation, we just execute \inysim on~$N_M$.
The following lemma (together with Lemma~\ref{lemma:iny}) implies the correctness of this approach.
%
% provided that $\inysim$ correctly computes the NFA simulation preorder.
%
%The following lemma shows that if $\simul$ is the simulation preorder
%over~$N_M$, then it is also the simulation preorder over~$M$.
%\ol{put the transformation from the lemma into prelims?}

\begin{lemma}\label{lemma:globalMinterms}
%  For an SFA $M = (Q, \algebra, \Delta, I, F)$ and its syntactic NFA $N = (Q, \alphabetof \Delta, \Delta, I, F)$, 
%if $M$  globally mintermised, then ${\simul_M} = {\simul_N}$.
  %Then $\inysim((Q, \alphabetof \Delta, \Delta, I, F))$ returns $\simul_M$.
Let $N = (Q, \alphabetof \Delta, \Delta, I, F)$ be the syntactic NFA of a
globally mintermised SFA $M = (Q, \algebra, \Delta, I, F)$. 
Then ${\simul_M} = {\simul_N}$.
\end{lemma}

The lemma below shows the time complexity of
\globalsim in terms of $n = |Q|$, $m = |\Delta|$, 
and 
the size $k$ of
the~largest~predicate~used~in~$\Delta$.

\begin{lemma}\label{lemma:globalsim} \globalsim computes $\simul_M$ in time
$\bigO{n m 2^{m} + \Csat(m,k) 2^m}.$ \end{lemma}
Intuitively, the complexity follows from the fact each transition of $\Delta$ can be
replaced by at most $2^m$ transitions in $\Delta_G$ since there can be
at most $2^m$ minterms in $\minterms{\Predicates_\Delta}$.
Nevertheless, $2^m$ minterms will always be generated (some of them
unsatisfiable, though), each of them generated from $m$~predicates of size at
most~$k$.
More details are available in Appendix~\ref{app:globalSimCompl}.

% \begin{wrapfigure}[22]{r}{7.8cm}
% \vspace*{-9mm}
% \hspace*{-3mm}
% \begin{minipage}{8.05cm}
%   \begin{algorithm}[H]
%     \DontPrintSemicolon
%
%     \KwIn{An SFA $M = (Q, \algebra, \Delta_M, I, F)$}
%     \KwOut{The simulation preorder ${}\preceq{} \subseteq Q \times Q$}
%     
%     compute global minterm normalized form of $M$, $\MG = (Q, \algebra, \Delta_{\MG}, I, F)$\;
%     
%     \lForAll{$q,p \in Q, \varphi \in \minterms{M}$} {
%       $N_{\varphi}(q,p) \asgn \card{\set{r | \markedmove(p,\varphi,r,\MG)}}$
% 	}
%
%     $\mathit{Rel} \asgn {(Q \times Q) \setminus (F \times (Q \setminus F))}$\;
%     $\mathit{NotRel} \asgn F \times (Q \setminus F)$\;
%
%
%     \While {$\mathit{NotRel} \neq \emptyset$} {
%       remove some $(i,j)$ from $\mathit{NotRel}$\;
%       \ForAll{$t \in Q, \varphi_{tj} \in \minterms{M}$ s.t. $\markedmove(t,\varphi_{tj},j,\MG)$} {
%         $N_{\varphi_{tj}}(i,t) \asgn N_{\varphi_{tj}}(i,t) - 1$\;
%         \If {$N_{\varphi_{tj}}(i,t) = 0$} {
%           \For {$s \in Q, \varphi_{si} \in \minterms{M}$ s.t. $\markedmove(s,\varphi_{si},i,\MG)$ and $(s,t) \in \mathit{Rel}$} {
%             $\mathit{Rel} \asgn {\mathit{Rel} \setminus \{(s,t)\}}$\;
%             $\mathit{NotRel} \asgn {\mathit{NotRel} \cup \{(s,t)\}}$\;
%           }
%         }
%       }
%     }
%     \Return $\mathit{Rel}$\;
%
%     \caption{GlobalSim}
%     \label{alg:GlobalSim}
%   \end{algorithm}
% \end{minipage}
% \end{wrapfigure}

%\newpage %!!!!!!!!!!!!!!!!!!

%*******************************************************************************
%\vspace{-0.0mm}
\subsection{Local Mintermisation-based Algorithm for SFAs (\localsim)}\label{sec:localsim}
%\vspace{-0.0mm}
%*******************************************************************************

Our next algorithm, called \localsim (Algorithm~\ref{alg:LocalSim}), represents
an attempt of running $\inysim$ on the original SFA without the global
mintermisation used above.
The main challenge in \localsim is how to symbolically represent the counters
$N_a(q,r)$---representing them explicitly would contradict the idea of symbolic
automata and would be impossible if the domain $\domain$~were infinite.
We will therefore use counters $N_\psi(q,r)$ indexed with labels $\psi$ of
outgoing transitions of $q$ to represent all counters $N_a(q,r)$, with $a\in\denote
\psi$.
A difficulty here is that if the automaton is not globally mintermised, then for
some $\move(q,\phi,p)$ and $a,b\in \denote\phi$, the sizes of $\goabove
q  a r$ and $\goabove q  b r$ may differ and hence cannot be represented by a
single counter.% 
\footnote{When describing an algorithm that works over an SFA, we use the
notation $\goabove q a r$ to represent the set
$\postofin{a}{q}{\denote{\Delta}}\cap\Rel(r)$, i.e., it refers to the
\emph{concrete} transitions of~$\denote{\Delta}$.}
For~example, if the only outgoing transition of $q$ other than $\move(q,\phi,p)$
is $\move(q,\psi,r)$ with $(p,r)\in\Rel$, $\denote \phi = \{a,b\}$, and $\denote{\psi} = \{b\}$, then
$|\goabove q  a r| = 1$ while $|\goabove q  b r| = 2$.
To~avoid this problem, we introduce the so-called local mintermised form,
in which only labels on outgoing transitions of every state must form
a~partition.

%\begin{wrapfigure}[17]{r}{76.0mm}
%\vspace*{-1mm}
%\hspace*{-3mm}
%\begin{minipage}{78.5mm}
  \begin{algorithm}[t]
    % \DontPrintSemicolon

    \KwIn{A complete SFA $M = (Q, \algebra, \Delta, I, F)$}
    \KwOut{The simulation preorder $\simul_M$}

    $\Delta_L \asgn $ locally mintermised form of $\Delta$\;

    \For{$p,q \in Q, \move(q,\psi,t) \in \transml$} { \label{algSFA:startcountinit}
      % \For{$\psi \in \minterms{q}$} {
      % \For{$\psi \in \minterms{q}$} {
        % \For{$p \in Q$} {
          $N_{\psi}(q,p) \asgn \card{\postofin \psi q {\Delta_L}}$ \label{algSFA:countinit}\;
          % $N_{\psi}(p,q) \asgn \card{\set{r | \markedmove(q,\psi,r,\ML)}}$\; \label{algSFA:countinit}
        % }
      % }
    } \label{algSFA:endcountinit}

    %$\Rel \asgn {(Q \times Q) \setminus (F \times (Q \setminus F))}$\; \label{algSFA:omega}
    %$\Rel \asgn {(Q \times Q)}$\;\label{algSFA:omega} 
    %$\NotRel \asgn F \times (Q \setminus F)$\; \label{algSFA:C}
    $\Rel \asgn {Q \times Q}$; 
    $\NotRel \asgn F \times (Q \setminus F)$ \label{algSFA:C} 
    
    \While {$\NotRel \neq \emptyset$} {
      remove some $(i,j)$ from $\NotRel$ and $\Rel$\; \label{algSFA:enq}
      \For{$\move(t, \psi_{tj}, j) \in \transml$} {
          $N_{\psi_{tj}}(t,i) \asgn N_{\psi_{tj}}(t,i) - 1$\; \label{algSFA:dec}
        \If(\tcp*[f]{$\goabove t {\psi_{tj}}i = \emptyset$}){$N_{\psi_{tj}}(t,i) = 0$} { \label{algSFA:zero}
          \For {$\move(s, \varphi_{si}, i) \in \Delta$ s.t. $(s,t) \in \Rel$} { \label{algSFA:forLastStart}
            \If{$\issat{\psi_{tj} \wedge \varphi_{si}}$} { \label{algSFA:check}
             % $\Rel \asgn \Rel \setminus \{(s,t)\}$\;
              $\NotRel \asgn \NotRel \cup \{(s,t)\}$\; 
            } \label{algSFA:forLastEnd}
          }
        }
      }
    }

    \Return $\Rel$\;

    \caption{\localsim}
    \label{alg:LocalSim}
  \end{algorithm}
%\end{minipage}
%\end{wrapfigure}

Formally, we say that
an SFA $M = (Q, \algebra, \Delta, I, F)$ is \emph{locally mintermised} if for
every state $p\in Q$, the set $\Predicates_{\Delta,p} \eqdef
\{\phi\in\Predicates\mid \exists q: \move(p,\phi,q)\in\Delta\}$ of the
predicates used on the transitions starting from $p$ is a partition.
A locally mintermised form is obtained by replacing every transition $\move(p,\phi,q)$ by the set of transitions 
% $\{\move(p,\minterm,q)\mid \minterm \in \minterms{\Predicates_{\Delta,p}} \land \issat{\neg \minterm \lor \phi}\}$.
% make it the same style as for global mintermisation
$\{\move(p,\minterm,q)\mid \minterm \in \minterms{\Predicates_{\Delta,p}} \land \issat{\minterm \land \phi}\}$.
Local mintermisation can hence be considerably cheaper than global mintermisation 
as it is only exponential to the maximum out-degree of a state (instead of the number of transitions of the whole SFA).
The key property of a locally mintermised SFA $M_L$ is the following: for any
transition $\move(q,\phi,p)$ of~$M_L$ and a state $r\in Q$,
and for any value of $\Rel$,
it~holds that $|\goabove q  a r|$ is the same for all $a\in\denote{\phi}$. 
This means that the set of counters $\{N_a(q,r) \mid a\in\denote{\phi}\}$ for
all symbols in the semantics of~$\phi$ can be represented by a~single
counter~$N_\phi(q,r)$.

The use of only locally mintermised transitions also necessitates a~modification of the \textbf{for} loop on line~\ref{algFA:itersymbols} of \inysim.
In particular, the test on line~\ref{algFA:forLastStart} of \inysim,
which determines the states~$s$ that cannot simulate~$t$ over the
symbol~$a$, only checks syntactic equivalence of the symbols.
This could lead to incorrect results because (syntactically) different local
minterms of different source states $t$ and $s$ can still have overlapping
semantics.
It can, in particular, happen that if a~counter~$N_{\psi_{tj}}(t,
i)$, for some predicate~$\psi_{tj}$, reaches zero on line~\ref{algSFA:zero} of
\localsim, there is a~transition from
state~$s$ to~$i$ over a predicate~$\varphi_{si}$ different from $\psi_{tj}$ but with some symbol $a\in\denote{\varphi_{si}}\cap \denote{\psi_{tj}}$. 
Because of~$a$, the state $t$ cannot simulate~$s$, 
but this would not happen if the two predicates were only compared
syntactically.
\localsim solves this issue on lines~\ref{algSFA:forLastStart} and~\ref{algSFA:check},
where it iterates over all transitions entering~$i$ and leaving a~state~$s$ simulated by~$t$ (wrt~$\Rel$),
and tests whether the predicate $\varphi_{si}$ on the transition semantically intersects with~$\psi_{tj}$.

\localsim is correct only if the input SFA is complete.
As mentioned in~\sct{sec:prelims}, this is, however, not an issue,
since completion of an SFA is, unlike for NFAs, straightforward, and its cost
is negligible compared with the complexity of \localsim presented below.

%Let $n,m$, and $k$ be as in Lemma~\ref{lemma:globalsim},
%$m_q$ for $q \in Q$
%denote the \emph{out-degree} (i.e.\ the number of transitions leaving) state~$q$, and $\maxoutdeg =
%\text{max}\{ m_q \mid q \in Q \}$ is
%the overall \emph{maximum out-degree}. Then the complexity of \localsim can be
%characterised as follows. 

% !!!!! Spacing reduced to make Springer happy.

The lemma below shows the time complexity of
\localsim in terms of $n = |Q|$, $m = |\Delta|$, 
the size $k$ of
the~largest~predicate~used~in~$\Delta$,
the \emph{out-degree} $m_q$ for each $q \in Q$
(i.e.\ the number of transitions leaving $q$), 
and the overall \emph{maximum out-degree} $\maxoutdeg =
\text{max}\{ m_q \mid q \in Q \}$.

\begin{lemma}\label{lemma:localsim}
  \localsim derives $\simul_M$ in time
  $$\bigO{n \sum_{q \in Q} m_q2^{m_q} +
  m \Csat(\maxoutdeg,k) \sum_{q \in Q} 2^{m_q}}.$$
\end{lemma}
As shown in more detail in Appendix~\ref{app:localSimCompl}, the result can be
proved in a similar way as in the case of \inysim and \globalsim, taking into
account that each transition is, again, replaced by its mintermised versions. 
This time, however, the mintermised versions are computed independently and
locally for each state (and the complexities are summed). 
Consequently, the factor $2^m$ gets replaced by $2^{m_q}$ for the different
states $q \in Q$ (together with the replacement of $\Csat(m,k)$ by
$\Csat(\maxoutdeg,k)$), which can significantly decrease the complexity. On the other
hand, as mintermisation is done separately for each state (which can sometimes
lead to re-doing some work done only once in \globalsim) and as one needs the
satisfiability test on line \ref{algSFA:check} of \localsim instead of the
purely syntactic test on line \ref{algFA:forLastStart} of \inysim, on which
\globalsim is based, \globalsim can sometimes win in practice.
This fact shows up even in our experiments presented
in~\sct{sec:experiments}.

%!!!!!!!!!!!!!!!!!!!!!!!!!!!!!!!!
%\enlargethispage{3mm}
%!!!!!!!!!!!!!!!!!!!!!!!!!!!!!!!!

%*******************************************************************************
%\vspace{-3.0mm}
\subsection{Counter-Free Algorithm for SFAs (\nocountsim)}\label{sec:nocount}
%\vspace{-1.0mm}
%*******************************************************************************

%Before we state our last algorithm, named \nocountsim
%(Algorithm~\ref{alg:NoCountSim}), 
%we introduce some notation.
%%
%Given an SFA $M = (Q, \algebra, \Delta, I, F)$, a~set $S\subseteq Q$,
%and a~state $q\in Q$, we use $\Reach{q}{S}$ denotes the disjunction of all predicates that
%reach $S$ from~$q$.
%Further, we write $\moves{q}{S}$ to denote that there exists a transition
%from~$q$ to some state in $S$.

Before we state our last algorithm, named \nocountsim
(Algorithm~\ref{alg:NoCountSim}), 
let us recall that
given an SFA $M = (Q, \algebra, \Delta, I, F)$, a~set $S\subseteq Q$,
and a~state $q\in Q$, we use $\Reach{q}{S}$ to denote the disjunction of all predicates that
reach $S$ from~$q$.
We will also write $\moves{q}{S}$ to denote that there is a transition
from~$q$ \mbox{to some state in $S$.}

%\begin{wrapfigure}[15]{r}{71mm}
%\vspace*{-9mm}
%\hspace*{-3mm}
%\begin{minipage}{73mm}
  \begin{algorithm}[t]
    % \DontPrintSemicolon

    \KwIn{A complete SFA $M = (Q, \algebra, \Delta, I, F)$}
    \KwOut{The simulation preorder $\simul_M$}

    %$\Rel \asgn {(Q \times Q) \setminus (F \times (Q \setminus F))}$ \label{algNoCountSFA:Rel:init}
    $\Rel \asgn {Q \times Q}$;$\NotRel \asgn F \times (Q \setminus F)$\;  \label{algNoCountSFA:Rel:init}

    %\lFor{$q \in F$}{$\NotRel_q \asgn Q \setminus F$}
    
    %\lFor{$q \in Q \setminus F$}{$\NotRel_q \asgn \emptyset$} \label{algNoCountSFA:endInit}

    \While {$\exists i \in Q: \NotRel(i) \neq \emptyset$} {

      $\Rm \asgn \{ t \mid \moves{t}{\NotRel(i)}\}$\; \label{algNoCountSFA:Rm}

      $\Rel(i)\asgn\Rel(i)\setminus\NotRel(i)$\;\label{algNoCountSFA:Rel}
		$\NotRel(i) \asgn \emptyset$\; \label{algNoCountSFA:NotRel}

      \For{$t \in \Rm$\label{algNoCountSFA:t}} {

        $\psi\asgn \Reach{t}{\Rel(i)}$\; \label{algNoCountSFA:psi}

          \For {$\move(s, \varphi_{si}, i) {\in} \Delta$ s.t. $(s,t)\in\Rel$} {\label{algNoCountSFA:sat}
            \If{$\issat{\neg\psi\land\varphi_{si}}$}{ \label{algAbsSFA:check} %\label{algSFA:check}
            %$\Rel \asgn \Rel \setminus \{(s,t)\}$\; \label{algNoCountSFA:Rel:update}
            %$\hspace{-2mm}\NotRel(s) {\asgn}\NotRel(s){\cup} \{t\}$\; \label{algNoCountSFA:NotRel:update}
            $\NotRel{\asgn}\NotRel{\cup} \{(s,t)\}$\; \label{algNoCountSFA:NotRel:update}
          } %\label{algNoCountSFA:secLoopEnd}
          }
        }
    }
    \Return $\Rel$\;

    \caption{\nocountsim}
    \label{alg:NoCountSim}
  \end{algorithm}
%\end{minipage}
%\end{wrapfigure}

In \nocountsim, we sacrifice the counting technique in order to avoid the local
mintermisation (which is still a~relatively expensive operation).
The obvious price for dropping the counters and local mintermisation is that the
emptiness of $\goabove t  a i$ for symbols $a\in\psi_{ti}$ can no more be tested
in a~constant time by asking whether ${N_{\psi_{ti}}(t,i) = 0}$ as on
line~\ref{algSFA:zero} of \localsim.
It does not even hold any more that $\goabove t a i$ is uniformly empty or
non-empty for all $a\in\psi_{ti}$.
To resolve the issue, we replace the test from line~\ref{algSFA:zero} of
\localsim by computing the formula $\psi =\Reach{t}{\Rel(i)}$ on
line~\ref{algNoCountSFA:psi} of \nocountsim, which is then used in the test on
line~\ref{algAbsSFA:check}.
Intuitively, $\psi$~represents all~$b$'s such that $\goabove t   b i$ is
\emph{not} empty.
By taking the negation of $\psi$, the test on line~\ref{algAbsSFA:check} of
\nocountsim then explicitly asks whether there is some $a \in
\denote{\varphi_{si}}$ for which $s$ can go to $i$ and $t$ cannot simulate
this move.

%In our third algorithm, named \nocountsim,
%we sacrifice the counting technique in order to 
%avoid the local mintermisation (which is still relatively expensive). 
%%
%The obvious price for dropping the counters and local mintermisation is that the emptiness of $\goabove t  a i$ for symbols $a\in\psi_{ti}$ can no more be tested (as on line~\ref{algSFA:zero} of \localsim) in a~constant time by asking whether ${N_{\psi_{ti}}(t,i) = 0}$,
%and it does not even hold any more that $\goabove t  a i$ is uniformly empty or not for all $a\in\psi_{ti}$.
%%
%This could be patched by skipping line~\ref{algSFA:zero} of \localsim, computing a~predicate~$\phi$ representing all symbols $a\in\psi_{tj}$ such that $\goabove t  a i$ got empty by the removal of $(i,j)$ from $\Rel$, and using $\phi$ instead of $\psi_{tj}$ on line~\ref{algSFA:check} of \localsim.
%%
%The predicate $\phi$ may be computed by (1)~constructing a~formula $\psi$ representing all~$b$'s such that $\goabove t   b i$ is \emph{not} empty---this is indeed implemented in \nocountsim by the computation of $\Reach{t}{\Rel(i)}$; recall 
%the convention that $\Rel(i)$ denotes the set of all $j$'s such that $(i,j)\in \Rel$ and $\Reach{t}{\Rel(i)}$ thus
%denotes the set of all symbols on transitions from $t$ to such~$j$'s---and
%(2) conjoining its negation with $\psi_{tj}$, that is, $\phi = \psi_{tj} \land
%\neg\psi$.
%%
%\ol{the previous is confusing---why doesn't it correspond to the algo?}

Further, notice that \nocountsim uses the set $\Rm$ for the following
optimisation.
Namely, if the use of $\Rm$ were replaced by an analogy of line~\ref{algSFA:enq}
from \localsim, it could choose a~sequence of several $j\in Q$ such that
$(i,j)\in\NotRel$, and then the same~$\psi$ would be constructed for each~$j$
%
% in the corresponding iteration of the \textbf{while} loop
%
and tested against the same~$\phi_{si}$.
%
% on line~\ref{algSFA:check} of \localsim.
%
In contrast, due to its use of $\Rm$, \nocountsim will process all
$j\in \NotRel(i)$ in a~single iteration of the main \textbf{while} loop, in
which $\psi$ is computed and tested against $\phi_{si}$ only once.

Lemma~\ref{lemma:nocountsim} shows the complexity of
\nocountsim in the terms used in Lemma~\ref{lemma:localsim}.  
\begin{lemma}\label{lemma:nocountsim}\nocountsim computes $\simul_M$ in time
  $\bigO{n\sum_{q \in Q}m_q^2 + m^2\Csat(\maxoutdeg,k)}.$ \end{lemma}
Observe that $\sum_{q \in Q}m_q = m$
and  $\maxoutdeg \leq n$, so the above complexity
is bounded by $\bigO{m^2\Csat(n,k)}$.
Out-degrees are, however, typically small constants.

The lemma is proved in Appendix~\ref{app:nocountSimCompl}.
Compared with the time complexity of \localsim, we can see that, by sacrificing
the use of the counters, the complexity becomes quadratic in the number of
transitions (since the decrement of the counter on line \ref{algSFA:dec}
followed by the test of the counter being zero on line \ref{algSFA:zero} in
\localsim is replaced by the computation of $\Gamma$ on
line~\ref{algNoCountSFA:psi} combined
with the test on line \ref{algAbsSFA:check} in \nocountsim).
On the other hand,
since we completely avoid mintermisation, the $2^{m_q}$ factors are lowered
to at most $m$ \mbox{($m_q$ in the left-hand side term).}

The overall worst-case complexity of \nocountsim is thus clearly better than
those of \globalsim and \localsim. Moreover, as shown in~\sct{sec:experiments},
\nocountsim is also winning in most of our experiments.
Another advantage of avoiding mintermisation is that it often requires a~lot of
memory.
Consequently, \globalsim and \localsim can run out of memory before even
finishing the mintermisation, which is also witnessed in our experiments.
If $m_q$ is small for all $q \in Q$ and the predicates do not intersect
much, the number of generated minterms can, however, be rather small compared with the
number of transitions, and \localsim can in some cases win, as witnessed in our
experiments too.

%%%%%%%%%%%%%%%%%%%%%%%%%%%%%%%%%%%%%%%%%%%%%%%%%%%%%%%%%%%%%%%%%%%%%%%%%%%%%%%%
%\vspace{-3.0mm}
\section{Experimental Evaluation}\label{sec:experiments}
%\vspace{-2.0mm}
%%%%%%%%%%%%%%%%%%%%%%%%%%%%%%%%%%%%%%%%%%%%%%%%%%%%%%%%%%%%%%%%%%%%%%%%%%%%%%%%

%%%%%%%%%%%%%%%%%%%%%%%%%%
% SETTINGS OF THE PLOTS
%%%%%%%%%%%%%%%%%%%%%%%%%%
\newcommand{\resizetofit}[1]{\resizebox{.325\textwidth}{!}{#1}}
\newcommand{\ourmark}[0]{+}

We now present an experimental evaluation of the algorithms
from~\sct{sec:computing} implemented in the Symbolic Automata
Toolkit~\cite{Veanes:2012:SAT:2260516.2260560}.
All experiments were run on an~Intel Core i5-3230M CPU@2.6\,GHz with 8\,GiB of
RAM.
We used the following two benchmarks:

%-----------------------------------------------------------------
\paragraph{\regex.}
We evaluated the algorithms on SFAs created from 1,921 regular expressions over
the UTF-16 alphabet using the $\BDD{16}$ algebra, which is the algebra of
\emph{binary decision diagrams} over 16 Boolean variables representing
particular bits of the UTF-16 encoding.
These regular expressions were taken from the website~\cite{RegexLib}, which
contains a~library of regular expressions %contributed by people all over the world,
created for different purposes, such as matching email addresses, URIs, dates,
times, street addresses, phone numbers, etc.
The SFAs created from these regular expressions were used before when evaluating
algorithms minimising (deterministic)
SFAs~\cite{DBLP:conf/popl/DAntoniV14} and when evaluating bisimulation
algorithms for SFAs~\cite{DAntoni2017}.
The largest automaton has 3,190 states and 10,702 transitions; the average
transition density of the SFAs is 2.5\,transitions per state.
Since the UTF-16 alphabet is quite large, a~symbolic representation is
needed for efficient manipulation of these automata.

%-----------------------------------------------------------------
\paragraph{\wsones.}
For this benchmark, we used 131 SFAs generated when deciding formulae of the
\emph{weak-monadic second order logic of one successor} (WS1S)~\cite{tata2007}.
We used two batches of SFAs: 93 deterministic ones from the tool
\mona~\cite{Elgaard:1998:MNT:647767.733623} and 38 nondeterministic from
\dwina~\cite{Fiedor:2015:NAW:2945565.2945584}.
These automata have at most 2,508 states and 34,374 transitions with the average
transition density of 6 transitions per state.
These SFAs use the algebra $\BDD{k}$ where $k$~is the number of variables in the
corresponding formula.

%*******************************************************************************
%\vspace{-0.0mm}
%\subsection{Evaluation}\label{sec:label}
\subsection{Comparison of Various Algorithms for Computing
Simulation}\label{sec:SimComp}
%\vspace{-0.0mm}
%*******************************************************************************

\begin{figure}[t]
\centering
% \begin{tikzpicture}
% \begin{axis}[
%     xlabel=global mint + RT,
%     ylabel=\localsim,
%     ymode=log,
%     xmode=log,
%     axis equal image,
%     ymin=1,
%     xmin=1,
%     ymax=100000,
%     xmax=100000,
%     width=\graphwidth,
% ]
% %\addplot+[mark size=1, only marks] table [x=globsimav, y=locsimoptav, col sep=comma] {data/final/regexnormalsupertrans.csv};
% \addplot+[mark size=1, only marks] table [x=globalSFA, y=localSFA, col sep=comma] {data/onlythisisused/comparison/regex.csv};
% \addplot+[mark=none,color=black] plot coordinates {(1,1) (100000,100000)};
% \end{axis}
% \end{tikzpicture}
\resizetofit{%
\begin{tikzpicture}
\begin{axis}[
    xlabel=\globalsim (39),
    ylabel=\localsim (98),
    ymode=log,
    xmode=log,
    axis equal image,
    ymin=1,
    xmin=1,
    ymax=10000,
    xmax=10000,
    width=\graphwidth,
]
%\addplot+[mark size=1, only marks] table [x=globsimav, y=locsimoptav, col sep=comma] {data/final/regexnormalsupertrans.csv};
\addplot+[mark size=1, only marks,mark=\ourmark] table [x=INYglobalSFA, y=localSFA, col sep=comma] {data/onlythisisused/comparison/regex2.csv};
\addplot+[mark=none,color=black] plot coordinates {(1,1) (10000,10000)};
\end{axis}
\end{tikzpicture}
}
% \begin{tikzpicture}
% \begin{axis}[
%     xlabel=\globalsim,
%     ylabel=\nocountsim,
%     ymode=log,
%     xmode=log,
%     axis equal image,
%     ymin=1,
%     xmin=1,
%     ymax=100000,
%     xmax=100000,
%     width=\graphwidth,
% ]
% %\addplot+[mark size=1, only marks] table [x=globsimav, y=nocountbettersimav, col sep=comma] {data/final/regexnormalsupertrans.csv};
% \addplot+[mark size=1, only marks] table [x=INYglobalSFA, y=nocountSFA, col sep=comma] {data/onlythisisused/comparison/regex.csv};
% \addplot+[mark=none,color=black] plot coordinates {(1,1) (100000,100000)};
% \end{axis}
% \end{tikzpicture}
\resizetofit{%
\begin{tikzpicture}
\begin{axis}[
    xlabel=\nocountsim (128),
    ylabel=\localsim (9),
    ymode=log,
    xmode=log,
    axis equal image,
    ymin=1,
    xmin=1,
    ymax=10000,
    xmax=10000,
    width=\graphwidth,
]
%\addplot+[mark size=1, only marks] table [x=nocountbettersimav, y=locsimoptav, col sep=comma] {data/final/regexnormalsupertrans.csv};
\addplot+[mark size=1, only marks,mark=\ourmark] table [x=nocountSFA, y=localSFA, col sep=comma] {data/onlythisisused/comparison/regex2.csv};
\addplot+[mark=none,color=black] plot coordinates {(1,1) (10000,10000)};
\end{axis}
\end{tikzpicture}
}
\resizetofit{%
\begin{tikzpicture}
\begin{axis}[
    xlabel=\gmrtsim (82),
    ylabel=\nocountsim (55),
    ymode=log,
    xmode=log,
    axis equal image,
    ymin=1,
    xmin=1,
    ymax=10000,
    xmax=10000,
    width=\graphwidth,
]
%\addplot+[mark size=1, only marks] table [x=globsimav, y=nocountbettersimav, col sep=comma] {data/final/regexnormalsupertrans.csv};
\addplot+[mark size=1, only marks,mark=\ourmark] table [x=globalSFA, y=nocountSFA, col sep=comma] {data/onlythisisused/comparison/regex2.csv};
\addplot+[mark=none,color=black] plot coordinates {(1,1) (10000,10000)};
\end{axis}
\end{tikzpicture}
}
\caption{Comparison of runtimes of algorithms on SFAs from \regex. Times are in miliseconds  (logarithmic scale).}
\label{fig:regex}
%\vspace{-5mm}
\end{figure}
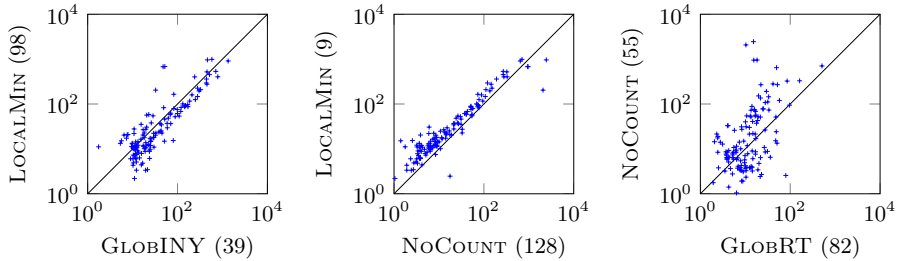

We first evaluate the effect of our modifications of \inysim presented
in~\sct{sec:computing}.
The results presented below clearly show the superiority of our new algorithms over
\globalsim, with \nocountsim being the overall winner.
In addition, we also compare the performance of our new algorithms to a version
of the RT algorithm from~\cite{RT}, which is one of the
best simulation algorithms.
In particular, we use its adaptation for NFAs, which we run after global
mintermisation (similarly as \inysim in \globalsim).
We denote the whole combination \gmrtsim.
\rtsim~is much faster than \inysim due to its use of the so-called
partition-relation pairs to represent the intermediate preorder.  Its \Cpp{}
implementation in the  \vata library~\cite{LengalSV2012} is also much more
optimised than the \Csharp{} implementation of our algorithms.  
Despite that, the comparison on automata with many global minterms is clearly
favourable to our new algorithms.

%!!!!!!!!!!!!!!!!!!!!!!!!!!!!!!!!
%\enlargethispage{3mm}
%!!!!!!!!!!!!!!!!!!!!!!!!!!!!!!!!

%we also compare the
%In addition to that, we also compare our new algorithms to the modification \cite{} of \globalsim \cite{} that globally mintermises the
%input SFA and then uses an implementation of the partition-based algorithm
%\rtsim~\cite{RT} in the \vata library~\cite{LengalSV2012} on the result
%(denoted as \gmrtsim).

%
%\tv{Somewhere, not necessarily here (perhaps in the intro?), we should explain
%that we have so far not used RT since it is not simple to apply it over SFAs but
%it is for sure an interesting subject for future research. Otherwise, the
%question why we have not gone for SFAs will naturally show up.}

%------------------------------------------------------------------------------
%\paragraph{Performance of the algorithms.}

% In the first experiment, we compared the run times of the algorithms.

%\tv{I do not like that we first characterize the results and only then give the
%concrete results. It would be better to start with the experimental results and
%then conclude with the above. I tried, but, I'm unfortunately too tired and I
%didn't manage. So I just rephrased the stuff a bit.}

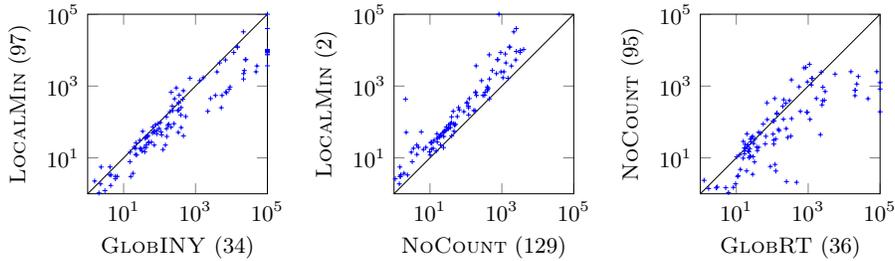
\begin{figure}[t]
\centering
%\includestandalone{time-WS1S-LoGlo}
% \begin{tikzpicture}
% \begin{axis}[
%     xlabel=global mint,
%     ylabel=local mint,
%     ymode=log,
%     xmode=log,
%     axis equal image,
%     ymin=1,
%     xmin=1,
%     ymax=100000,
%     xmax=100000,
%     width=\graphwidth,
% ]
% %\addplot+[mark size=1, only marks] table [x=globsim, y=locsimopt, col sep=comma] {data/final/timbuktnormaltrans.csv};
% \addplot+[mark size=1, only marks] table [x=globalSFA, y=localSFA, col sep=comma] {data/onlythisisused/comparison/ws1s.csv};
% \addplot+[mark=none,color=black] plot coordinates {(1,1) (100000,100000)};
% \end{axis}
% \end{tikzpicture}
\resizetofit{%
%\includestandalone{time-WS1S-LoNo}
\begin{tikzpicture}
\begin{axis}[
    xlabel=\globalsim (34),
    ylabel=\localsim (97),
    ymode=log,
    xmode=log,
    axis equal image,
    ymin=1,
    xmin=1,
    ymax=100000,
    xmax=100000,
    width=\graphwidth,
]
%\addplot+[mark size=1, only marks] table [x=globsim, y=locsimopt, col sep=comma] {data/final/timbuktnormaltrans.csv};
\addplot+[mark size=1, only marks,mark=\ourmark] table [x=INYglobalSFA, y=localSFA, col sep=comma] {data/onlythisisused/comparison/ws1s2.csv};
\addplot+[mark=none,color=black] plot coordinates {(1,1) (100000,100000)};
\end{axis}
\end{tikzpicture}
}
%\includestandalone{time-WS1S-NoGlo}
% \begin{tikzpicture}
% \begin{axis}[
%     xlabel=INYglobal mint,
%     ylabel=nocount,
%     ymode=log,
%     xmode=log,
%     axis equal image,
%     ymin=1,
%     xmin=1,
%     ymax=100000,
%     xmax=100000,
%     width=\graphwidth,
% ]
% %\addplot+[mark size=1, only marks] table [x=globsim, y=nocountbettersim, col sep=comma] {data/final/timbuktnormaltrans.csv};
% \addplot+[mark size=1, only marks] table [x=INYglobalSFA, y=nocountSFA, col sep=comma] {data/onlythisisused/comparison/ws1s.csv};
% \addplot+[mark=none,color=black] plot coordinates {(1,1) (100000,100000)};
% \end{axis}
% \end{tikzpicture}
\resizetofit{%
\begin{tikzpicture}
\begin{axis}[
    xlabel=\nocountsim (129),
    ylabel=\localsim (2),
    ymode=log,
    xmode=log,
    axis equal image,
    ymin=1,
    xmin=1,
    ymax=100000,
    xmax=100000,
    width=\graphwidth,
]
%\addplot+[mark size=1, only marks] table [x=nocountbettersim, y=locsimopt, col sep=comma] {data/final/timbuktnormaltrans.csv};
\addplot+[mark size=1, only marks,mark=\ourmark] table [x=nocountSFA, y=localSFA, col sep=comma] {data/onlythisisused/comparison/ws1s2.csv};
\addplot+[mark=none,color=black] plot coordinates {(1,1) (100000,100000)};
\end{axis}
\end{tikzpicture}
}
\resizetofit{%
%\includestandalone{time-WS1S-NoGlo}
\begin{tikzpicture}
\begin{axis}[
    xlabel=\gmrtsim (36),
    ylabel=\nocountsim (95),
    ymode=log,
    xmode=log,
    axis equal image,
    ymin=1,
    xmin=1,
    ymax=100000,
    xmax=100000,
    width=\graphwidth,
]
%\addplot+[mark size=1, only marks] table [x=globsim, y=nocountbettersim, col sep=comma] {data/final/timbuktnormaltrans.csv};
\addplot+[mark size=1, only marks,mark=\ourmark] table [x=globalSFA, y=nocountSFA, col sep=comma] {data/onlythisisused/comparison/ws1s2.csv};
\addplot+[mark=none,color=black] plot coordinates {(1,1) (100000,100000)};
\end{axis}
\end{tikzpicture}
}
\caption{Comparison of runtimes of algorithms on SFAs from \wsones. Times are in miliseconds (logarithmic scale).}
\label{fig:ws1s}
%\vspace{-0mm}
\end{figure}

To  proceed to concrete data, Figs.~\ref{fig:regex} and~\ref{fig:ws1s} show scatter plots
of the most interesting comparisons of the runtimes of the considered
algorithms on our benchmarks (we give in parentheses the number of times the
corresponding algorithm \emph{won} over the other one).
The timeout was set to 100\,s.
Fig.~\ref{fig:regex} shows the comparison of the algorithms on SFAs from the
\regex benchmark.
In this experiment, we removed the SFAs where all algorithms finished within 10\,ms
(to mitigate the effect of imprecise measurement and noise caused by the
\Csharp{} runtime), which gave us 138~SFAs.
Moreover, we also removed one extremely challenging SFA, which dominated the
whole benchmark (we report on that SFA, denoted as $M_c$, later), which
left us with the final number of 137 SFAs.
On the other hand, Fig.~\ref{fig:ws1s} shows the comparison for \wsones.
We observe the following phenomena:
\begin{inparaenum}[(i)]
  \item  \localsim is in the majority of cases faster than \globalsim,
  \item  \nocountsim clearly dominates \localsim, and
  \item  the comparison of \nocountsim and \gmrtsim has no clear winner: on the
    \regex benchmark, \gmrtsim is more often faster, but on the \wsones
    benchmark, \nocountsim wins
    (in many cases, quite significantly).
\end{inparaenum}

Further, we also give aggregated results of the experiment in
Table~\ref{tab:times}.
In the table, we accumulated the runtimes of the algorithms over the whole
benchmark (column ``time'') and the number of times each algorithm was the best
among all algorithms (column ``wins'').
The column ``fails'' shows how many times the
respective algorithm failed (by being out of time or memory).
In the parentheses, we give the number of times the failure occurred already in
the mintermisation.
When a benchmark fails, we assign it the time 100\,s (the timeout)
for the computation of ``time''.
The times of the challenging SFA $M_c$ from \regex were:
21\,s for \globalsim,
16\,s for \localsim,
25\,s for \nocountsim, and
148\,s for \gmrtsim.
Obviously, including those times would bias the whole evaluation.

%\begin{wraptable}[9]{r}{6.25cm}
\begin{table}[t]
%  \vspace*{-8mm}
%  \vspace*{-3mm}
%  \hspace*{-2mm}
%  \begin{minipage}{6.4cm}
\caption{Aggregated results of the performance experiment.}
  \centering
  {%\scriptsize
\begin{tabular}{|l||r|r|r|r|r|}
  \hline
                & \multicolumn{2}{c|}{\regex}  & \multicolumn{3}{c|}{\wsones}  \\
  Algorithm     & \multicolumn{1}{c|}{time}    & \multicolumn{1}{c|}{wins}  & \multicolumn{1}{c|}{time}     & \multicolumn{1}{c|}{wins} & \multicolumn{1}{c|}{fails}       \\ \hline \hline
  \globalsim    & 12.3\,s  & 2      & 1,258\,s  & 1    & 9 (2)      \\
  \localsim     & 11.9\,s  & 0      &   316\,s  & 0    & 1 (1)      \\
  \nocountsim   & 12.4\,s  & 54     &    44\,s  & 94   & 0 (0)      \\
  \gmrtsim      &  2.8\,s  & 81     &   594\,s  & 36   & 3 (2)      \\
  \hline
\end{tabular}
}
\vspace{2mm}
%  \end{minipage}
\label{tab:times}
\end{table}
%\end{wraptable}

Observe that in this comparison, the performance of the algorithms on the two
benchmarks differs---although \gmrtsim wins on the \regex
benchmark and the other three algorithms have a~comparable overall time (but
\nocountsim still wins in the majority of SFAs among the three), on the more
complex benchmark (\wsones), \nocountsim is the clear winner.
% \lh{I think we should interpret the results differently. In the table, we should sum only non-timeouts while the one regex outlayer of RT (148sec) should be a timeout (it indeed runs more than 100 seconds). The RT will then totally win on the regexes (with the overall runtime about 3s), up to the one timeout. Would be good if we could differentiate the two classes of benchmarks based on the number of minterms or something like that. Overall, we cannot really say that nocount wins over RT. We have to change the story a bit.}
The distinct results on the two benchmark sets can be explained by a~different
diversity of predicates used on the transitions of SFA.
In the \regex benchmark, the globally mintermised automaton has on average 4.5
times more transitions (with the ratio ranging from 1 to 13), while in the WS1S
benchmark, the mintermised automaton has on average 23.5 times more transitions
(with the ratio ranging from 1 to 716).
This clearly shows that our algorithms are effective in avoiding the potential
blow-up of mintermisation.
As expected, they are slower than \rtsim on examples where the mintermisation is
cheap since they do not use the \mbox{partition-relation data structure.}

%\begin{wrapfigure}[11]{r}{4.34cm}
%\vspace*{-8mm}
%\hspace*{-2mm}
%\begin{figure}[t!]
%\resizebox{4.5cm}{!}{
%\begin{tikzpicture}
%\begin{axis}[
%    xlabel=simulation,
%    ylabel=bisimulation,
%    ymode=log,
%    xmode=log,
%    axis equal image,
%    ymin=1,
%    xmin=1,
%    ymax=10000,
%    xmax=10000,
%    width=\graphwidthother,
%]
%\addplot+[mark size=1, only marks,mark=\ourmark] table [x=transitionsSim, y=transitionsBisim, col sep=comma] {data/onlythisisused/simvsbisimnew/Regex.csv};
%\addplot+[mark=none,color=black] plot coordinates {(1,1) (10000,10000)};
%\end{axis}
%\end{tikzpicture}
%}
%%\vspace{-8mm}
%\caption{Reduction of transitions.}
%\label{fig:sim_bisim}
%%\end{wrapfigure}
%\end{figure}

\begin{figure}[t]
 \centering
 \begin{subfigure}{0.45\textwidth}
 \centering
 \begin{tikzpicture}
%  \begin{tikzpicture}[scale = .75]
 \begin{axis}[
     xlabel=simulation,
     ylabel=bisimulation,
     axis equal image,
     ymin=0,
     xmin=0,
     ymode=log,
     xmode=log,
     width=\graphwidthother,
 ]
 \addplot+[mark size=1, only marks,mark=\ourmark] table [x=transitionsSim, y=transitionsBisim, col sep=comma] {data/onlythisisused/simvsbisim/Regex.csv};
 \addplot+[mark=none,color=black] plot coordinates {(1,1) (10000,10000)};
 \end{axis}
 \end{tikzpicture}
 \caption{Iterative reduction.}
 \label{fig:simbisim-transitions-iteration}
 \end{subfigure}
 \begin{subfigure}{0.45\textwidth}
 \centering
 \begin{tikzpicture}
%  \begin{tikzpicture}[scale=.75]
 \begin{axis}[
     xlabel=simulation,
     ylabel=bisimulation,
     axis equal image,
     ymin=0,
     xmin=0,
     ymode=log,
     xmode=log,
     width=\graphwidthother,
 ]
 \addplot+[mark size=1, only marks,mark=\ourmark] table [x=transitionsSim, y=transitionsBisim, col sep=comma] {data/onlythisisused/simvsbisim/onlyoneRegex.csv};
 \addplot+[mark=none,color=black] plot coordinates {(1,1) (10000,10000)};
 \end{axis}
 \end{tikzpicture}
 \caption{One iteration only.}
 \label{fig:simbisim-transitions-once}
 \end{subfigure}
 \caption{Simulation vs. bisimulation-based reduction: the number of transitions of the reduced automaton.}
\label{fig:simbisim-transitions}
\end{figure}
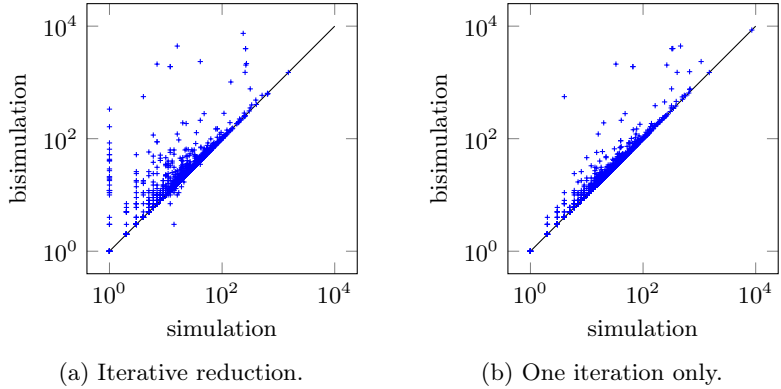

%------------------------------------------------------------------------------
%\paragraph{Comparison of simulation and bisimulation.}
\subsection{Comparison of Simulation and Bisimulation}

In the second experiment, we evaluate the benefit of computing simulation over
computing bisimulation (we use the implementation of bisimulation computation
from~\cite{DAntoni2017}).
In particular, we focus on an application of (bi-)simulation for
(language-preserving) reduction of SFAs from the whole \regex benchmark.

For every SFA~$M$ from the benchmark, we compute its simulation
preorder~$\simul_M$, take its biggest symmetric fragment (which constitutes an
equivalence), and for each of its classes, merge all states of the class into
a~single state.
%Because $\simul$ is a~preorder,
%\ol{why is that a reason?}
We also eliminate simulation subsumed 
transitions (the so-called \emph{little brothers}) using the technique introduced 
in~\cite{Bustan:2003:SM:635499.635502}.
In~particular, for a~state $q$ s.t.\
there exist transitions $\move(q,a,p)$ and $\move(q,a,p')$ with $p \simul p'$, we
remove the transition $\move(q,a,p)$ (and also the states that have
become unreachable).
After that, we reverse the automaton and repeat the whole procedure.
These steps continue until the number of states no longer decreases.
Similar steps apply to bisimulation (with the exception of taking the symmetric
fragment and removing transitions as a~bisimulation is already an equivalence).

The results comparing the number of transitions of the output SFAs are given in
Fig.~\ref{fig:simbisim-transitions-iteration}, showing that the simulation-based
reduction is usually much more significant.\footnote{%
There are still some cases when bisimulation achieved a larger
reduction than simulation, which may seem unintuitive since the largest
bisimulation is always contained in the simulation preorder.
This may happen, e.g., when a~simulation-based reduction disables an (even
greater) reduction on the subsequent reversed SFA.
}
 Fig.~\ref{fig:simbisim-transitions-once} shows the reduction after the first iteration (it corresponds to the ``ordinary'' simulation and bisimulation-based reduction).

The comparison of the numbers of states gives a very similar picture as the comparison of the numbers of transitions (cf.\ Appendix~\ref{app:exp}) but simulation wins by a slightly larger margin when comparing the numbers of transitions. This is probably due to the use of the removal of simulation-subsumed transitions, which does not have a meaningful counterpart when working with bisimulations.

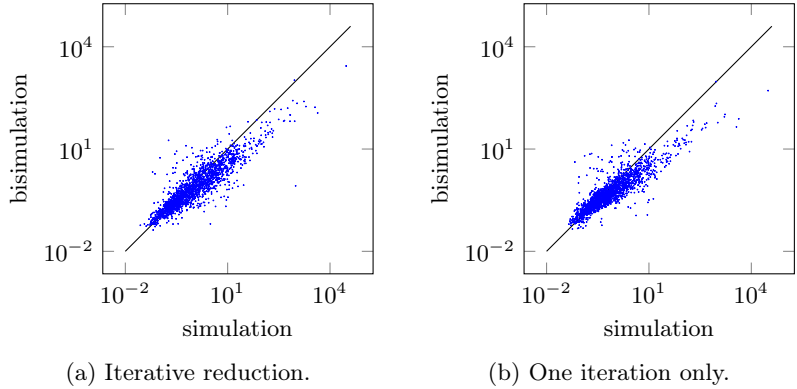
\begin{figure}[t]
 \centering
 \begin{subfigure}{0.45\textwidth}
 \centering
 \begin{tikzpicture}
%  \begin{tikzpicture}[scale = .75]
 \begin{axis}[
     scaled y ticks={base 10:-3},
     ytick scale label code/.code={},
     scaled x ticks={base 10:-3},
     xtick scale label code/.code={},
     xlabel=simulation,
     ylabel=bisimulation,
     axis equal image,
     ymin=0,
     xmin=0,
     ymode=log,
     xmode=log,
     width=\graphwidthother,
 ]
 \addplot+[mark size=0.3, only marks,mark=\ourmark] table [x=timeSim, y=timeBisim, col sep=comma] {data/onlythisisused/simvsbisim/Regex.csv};
 \addplot+[mark=none,color=black] plot coordinates {(0.01,0.01) (40000,40000)};
 \end{axis}
 \end{tikzpicture}
 \caption{Iterative reduction.}
 \label{fig:simbisim-time-iteration}
 \end{subfigure}
 \begin{subfigure}{0.45\textwidth}
 \centering
 \begin{tikzpicture}
%  \begin{tikzpicture}[scale = .75]
 \begin{axis}[
     scaled y ticks={base 10:-3},
     ytick scale label code/.code={},
     scaled x ticks={base 10:-3},
     xtick scale label code/.code={},
     xlabel=simulation,
     ylabel=bisimulation,
     axis equal image,
     ymin=0,
     xmin=0,
     ymode=log,
     xmode=log,
     width=\graphwidthother,
 ]
 \addplot+[mark size=0.3, only marks,mark=\ourmark] table [x=timeSim, y=timeBisim, col sep=comma] {data/onlythisisused/simvsbisim/onlyoneRegex.csv};
 \addplot+[mark=none,color=black] plot coordinates {(0.01,0.01) (40000,40000)};
 \end{axis}
 \end{tikzpicture}
 \caption{One iteration only.}
 \label{fig:simbisim-time-once}
 \end{subfigure}
 \caption{Simulation vs. bisimulation-based reduction: runtime in miliseconds. }
 \label{fig:simbisim-time}
 \end{figure}

As for the runtimes, they differ significantly on the different case studies
with some of the cases won by the simulation-based reduction process, some by
the bisimulation-based reduction, as can be seen in Fig.~\ref{fig:simbisim-time}.
Fig.~\ref{fig:simbisim-time-iteration} shows comparison of runtimes for the whole iterative  process, Fig.~\ref{fig:simbisim-time-once} shows the comparison for the first iteration only---essentially the time taken by computing the simulation preorder or the bisimulation equivalence.
%Additional details on this experiment are given in~Appendix~\ref{app:exp}.
%
%Lastly, Figure~\ref{fig:simbisim-time} shows a comparison of the running times of the simulation and bisimulation reduction. 
%Figure~\ref{fig:simbisim-time-iteration} shows the overall time needed by the iterative reduction process, Figure~\ref{fig:simbisim-time-once} then the time taken by the first iteration---essentially the time taken by computing the simulation preorder or the bisimulation equivalence. 
One may see that bisimulation is notably cheaper, especially when the automata are growing larger and both algorithms are taking more time (note the logarithmic scale). 
Computing simulation was, however, faster in surprisingly many cases.

%
% Surprisingly, the runtime of simulation is slightly better than the runtime of
% bisimulation (1.138 times on average)---however, the times are often comparable.
%
% \tv{What is this? Can one say x--y (on average z) times slower or something like that?}, 
% in many cases comparable, though.
%\ol{actually, the following was measured on the values (one run) of $\frac{time bisim}{time sim}$: mean = 1.138, sd = 4.123, max = 149.13, min = 0.000427 (we used \nocountsim for computing simulation)}
%\lh{the interesting thing is that the time needed for simulation is comparable with the time needed for bisimulation, sometimes smaller, somatimes bigger (details in the appendix). It is surprising that simulation is not always more expensive}.
% \ol{OLD:actually, the following was measured: on average, simulation is 3\,\% faster than bisim (one direction), with the standard deviation 239\,\%, maximum 6171\,\% and minimum 88\,\% slower.  Raw data: bisim/sim; mean = 1.037, sd = 2.385, max = 61.71, min = 0.0122 (we used \nocountsim for computing simulation)}
% Details of the experiment are given in~Appendix~\ref{app:exp}.
% \textcolor{red}{[XX]}

%%%%%%%%%%%%%%%%%%%%%%%%%%%%%%%%%%%%%%%%%%%%%%%%%%%%%%%%%%%%%%%%%%%%%%%%%%%%%%%%%
%\vspace{-0.0mm}
\section{Conclusion and Future Work}\label{sec:label}
%\vspace{-0.0mm}
%%%%%%%%%%%%%%%%%%%%%%%%%%%%%%%%%%%%%%%%%%%%%%%%%%%%%%%%%%%%%%%%%%%%%%%%%%%%%%%%%
%%
%%Here you are welcome to fill a bitcoin address where \$500 will be sent if this paper is accepted.   
%
We have introduced two new algorithms for computing simulation over symbolic
automata that do not depend on global mintermisation: one that needs a local
and cheaper variant of mintermisation, and one that does not need
mintermisation at all.
They perform well especially on automata where mintermisation significantly increases the number of transitions.
In the future, we would like to come up with a~partition-based algorithm that could run on an SFA without the need of mintermisation.
Such algorithm might, but does not necessarily need to, be based on an NFA
partition-based algorithm such as~\rtsim.
Further, we wish to explore the idea of encoding NFAs over finite alphabets
compactly as SFAs over a fast Boolean algebra (such as bit-vector encoding of
sets) and compare the performance of our algorithms with known NFA simulation
algorithms. 
%%

%\ol{do we want this section?}\tv{No need for conclusion but at least 1 sentence
%about future work would be nice. So perhaps 1 sentence that we presented 3
%algorithms for ... and presented their experimental comparison. In the future...}
%
%\ol{future work: extending to Ranzato \&
%Tapparo~\cite{DBLP:conf/lics/RanzatoT07}}\tv{Hopefully combining the orthogonal
%advantages of the approaches.}
%\blindtext

%------------------------------------------------------------------------------
\paragraph{Acknowledgements.}
The work on this paper was supported by
the Czech Science Foundation projects 16-17538S and 16-24707Y,
the IT4IXS: IT4Innovations Excellence in Science project (LQ1602),
and the FIT BUT internal project FIT-S-17-4014.

%%%%%%%%%%%%%%%%%%%%%%%%%%%%%%%%%%%%%%%%%%%%%%%%%%%%%%%%%%%%%%%%%%%%%%%%%%%%%%%%
% \bibliographystyle{splncs}
% \bibliography{literature}

%\vspace{-2mm}

%%%%%%%%%%%%%%%%%%%%%%%%%%%%%%%%%%%%%%%%%%%%%%%%%%%%%%%%%%%%%%%%%%%%%%%%%%%%%%%%

\newpage
\appendix

%%%%%%%%%%%%%%%%%%%%%%%%%%%%%%%%%%%%%%%%%%%%%%%%%%%%%%%%%%%%%%%%%%%%%%%%%%%%%%%%
\section{Complexity of the \inysim Algorithm}\label{app:inySimCompl}
%%%%%%%%%%%%%%%%%%%%%%%%%%%%%%%%%%%%%%%%%%%%%%%%%%%%%%%%%%%%%%%%%%%%%%%%%%%%%%%%

If $n = |Q|$ is the number of states, $m = |\Delta|$ is the number of
transitions, and $\ell = |\Sigma|$ is the size of the alphabet of an NFA $N =
(Q, \Sigma, \Delta, I, F)$, the time complexity of the \inysim algorithm is
$\bigO{nm+\ell n^2}$ as stated in Lemma~\ref{lemma:iny}.
As this fact is not immediately obvious, we give a proof of
Lemma~\ref{lemma:iny} below, building on~\cite{nfasim}.

\begin{proof}[Lemma~\ref{lemma:iny}] The initialization on
lines~\ref{algFA:startInit}--\ref{algFA:endInit} is done in $\bigO{m + \ell
n^2}$ time.
Since we save in $\NotRel$ pairs of states that are to be processed, and we save
each pair at most once, line~\ref{algFA:deq} is reached at most $n^2$ times.

Next, the value of the sum of the initial values of all counters can be
characterised as follows: \[\sum_{\substack{i,t \in Q \\ a \in \Sigma}}N_a(t,i)
= \sum_{\substack{i,t \in Q \\ a \in \Sigma}} \card{\postof a t}.\]
For a fixed transition $t \in Q$, the sum $\sum_{a \in \Sigma}\card{\postof a
t}$ is equal to the number of transitions going from the state $t$, and the sum
$\sum_{t \in Q}\sum_{a \in \Sigma}\card{\postof a t}$ is then equal to the
number of all transitions $m$.
Therefore, \[\sum_{\substack{i,t \in Q \\ a \in \Sigma}}\card{\postof a t} =
\sum_{i \in Q}\sum_{\substack{t \in Q \\ a \in \Sigma}}\card{\postof a t} =
\sum_{i \in Q}m = nm.\]
Also, since counters cannot be negative (because they represent the number of
states simulating some state), we can now say that line~\ref{algFA:dec}
(decrementing the counters) is reached at most $nm$ times.

Now, the only thing left to show is that
lines~\ref{algFA:forLastStart}--\ref{algFA:forLastEnd} are reached at most $nm$
times.
For that, we first note that if we fix $i \in Q, a \in \Sigma$ in $N_a(i,t)$,
line~\ref{algFA:forLastStart} is reached at most $n$ times (there are $n$ such
counters).
On the other hand, if we fix $t$, the \textbf{for} loop on
lines~\ref{algFA:forLastStart}--\ref{algFA:forLastEnd} is iterated at most $m$
times.
This stems from a similar fact as the argument for the initial sum of the
counters: the \textbf{for} loop enumerates all states $s$, such that
$\move(s,a,i)$, and summed over all states $i$ and symbols $a$, it computes its
body $m$ times (for a fixed $t$).
If we combine these two facts,
lines~\ref{algFA:forLastStart}--\ref{algFA:forLastEnd} are reached at most $nm$
times.

Overall, we showed that \inysim runs in $\bigO{nm+\ell n^2}$
time.\qed\end{proof}

%%%%%%%%%%%%%%%%%%%%%%%%%%%%%%%%%%%%%%%%%%%%%%%%%%%%%%%%%%%%%%%%%%%%%%%%%%%%%%%%
\section{Correctness and Complexity of the \globalsim
Algorithm}\label{app:globalSimCompl}
%%%%%%%%%%%%%%%%%%%%%%%%%%%%%%%%%%%%%%%%%%%%%%%%%%%%%%%%%%%%%%%%%%%%%%%%%%%%%%%%

In this appendix, we provide a proof of Lemma~\ref{lemma:globalMinterms},
underlying correctness of Algorithm \globalsim, and then provide a proof of
Lemma~\ref{lemma:globalsim}, stating the complexity of the algorithm.

\begin{proof}[Lemma~\ref{lemma:globalMinterms}] We will prove that $R \subseteq
{Q \times Q}$ is a~simulation on $M$ iff $R$ is a~simulation on $N$.

Let $R$ be a simulation on $M$ and assume that it is not simulation on $N$.
Then, there must be some $(q,p) \in R$ contradicting the definition of
simulation.
Since the sets of final states are the same, $(q,p)$ must contradict
Condition~C\ref{simFA2}.
This means that there is some $\move(q,\psi,q') \in \Delta$ for which there is
no $\move(p,\psi,p') \in \Delta$ such that $(q',p') \in R$.
However, since for all $a \in \domain_{\algebra}$ there exists exactly one
minterm $\phi$ for which $a \in \denote{\phi}$, this would mean that $(q,p)$
contradicts Condition~C\ref{simFA2} for $M$, which is a contradiction.

Since $M$ and $N$ have the same transition relation $\Delta$, the other
direction is obvious.\qed\end{proof}

\begin{proof}[Lemma~\ref{lemma:globalsim}] Apart from $n$ being the number of
states of $M$, $m$ being the number of its transitions, and $k$ being the size
of the largest predicate used in the transitions of $M$, let $m'$ be the number
of transition in $\Delta_G$.
Recall the definition of $\Csat$ from Section~\ref{sec:prelims}.
As there can be at most $2^m$ minterms in $\minterms{\Predicates_\Delta}$ and
every minterm is generated from $m$ predicates, we can compute them in
$\bigO{2^{m}\Csat(m,k)}$ time.
Computing $\Delta_G$ is then done in $\bigO{2^{m}\Csat(m,k) + m2^m}$: every
transition is replaced by transitions labelled with minterms.
As we then run \inysim on the syntactic NFA $(Q, \Predicates_{\Delta_G},
\Delta_G, I, F)$, we can conclude that Algorithm \globalsim has complexity
$\bigO{2^m\Csat(m,k) + m2^m + nm'}$.
Further, we can bound $m'$ by $m2^{m-1}$ because each transition can be at worst
replaced by $2^{m-1}$ transitions: indeed, note that the predicate of each
transition occurs in half of the minterms.
The complexity of Algorithm \globalsim is then \[\bigO{2^m\Csat(m,k) +
nm2^{m}}.\] \qed\end{proof}

%%%%%%%%%%%%%%%%%%%%%%%%%%%%%%%%%%%%%%%%%%%%%%%%%%%%%%%%%%%%%%%%%%%%%%%%%%%%%%%%
\section{Complexity of the \localsim Algorithm}\label{app:localSimCompl}
%%%%%%%%%%%%%%%%%%%%%%%%%%%%%%%%%%%%%%%%%%%%%%%%%%%%%%%%%%%%%%%%%%%%%%%%%%%%%%%%

In this appendix, we examine the time complexity of Algorithm \localsim and prove
Lemma~\ref{lemma:localsim}.

\begin{proof}[Lemma~\ref{lemma:localsim}] For a given state $q \in Q$, let $r_q$
be the number of minterms in $\minterms{\Predicates_{\Delta,q}}$ and $m_q$ the
number of transitions with the source state $q$.
Using the same reasoning as for global mintermisation, one can show that
$\Delta_L$ can be computed in time $\bigO{\sum_{q \in Q}(r_q\Csat(m_q,k) +
m_{q}r_q)}$.
Further, let $r$ be the number of all local minterms, i.e., $r = \sum_{q \in
Q}r_q$, and let $m'$ be the number of transitions in $\Delta_L$.
The initialization on lines~\ref{algSFA:startcountinit}--\ref{algSFA:C} is done
in $\bigO{nr}$ time: $\card{\set{r | \markedmove(q,\psi,r,\ML)}}$ is computed
during mintermisation.
Using the same reasoning as in NFA simulation, we can say that, initially, the
sum of all counters is $nm'$, and so line~\ref{algSFA:dec} is reached at most
$nm'$ times.
For a fixed $i$, line~\ref{algSFA:forLastStart} is reached $r$ times because
there are $r$ counters $N_{\psi_{tj}}(t,i)$ that can reach zero only once: for
each $t \in Q$ there is one counter $N_{\psi}(t,i)$ for each $\psi \in
\minterms{\Predicates_{\Delta,t}}$.
If we now fix $t$ and $\psi_{tj}$,
lines~\ref{algSFA:check}--\ref{algSFA:forLastEnd} are reached at most $m$ times.
All in all, these lines are reached at most $rm$ times.
Since $\psi_{tj}$ is a minterm created from $m_t$ transitions and $nr \leq nm'$,
the time complexity of the algorithm is $\bigO{\sum_{q \in Q}(r_q\Csat(m_q,k) +
m_{q}r_q) + nm' + mr\Csat(\maxoutdeg,k)}$.
Since, for a given $q \in Q$, $r_q$ is bounded by $2^{m_q}$ (because $r_q$ is
the number of minterms) and since $m'$ is bounded by $\sum_{q \in Q}m_q2^{m_q}$,
the final time complexity of Algorithm \localsim is \[\bigO{n\sum_{q \in
Q}m_q2^{m_q} + m\Csat(\maxoutdeg,k)\sum_{q \in Q}2^{m_q}}.\]\qed\end{proof}

%%%%%%%%%%%%%%%%%%%%%%%%%%%%%%%%%%%%%%%%%%%%%%%%%%%%%%%%%%%%%%%%%%%%%%%%%%%%%%%%
\section{Correctness and Complexity of the \nocountsim Algorithm}\label{app:nocountSimCompl}
%%%%%%%%%%%%%%%%%%%%%%%%%%%%%%%%%%%%%%%%%%%%%%%%%%%%%%%%%%%%%%%%%%%%%%%%%%%%%%%%

We first prove correctness of Algorithm \nocountsim, i.e., the fact that it
indeed computes $\simul_M$ on an SFA $M$.
Then, we establish the complexity of the algorithm stated in
Lemma~\ref{lemma:nocountsim}.

\subsection{Correctness}

We prove that \nocountsim computes $\simul_M$.  
We first prove that the following invariant is preserved by the algorithm for any
simulation relation $\simul$ over $Q$.
\[
\NotRel \subseteq \compl{\Rel} \subseteq {\compl{\simul}}
\]
Initially $\NotRel=\compl{\Rel}=F\times (Q\setminus F)$, and for all
$x\in F$ and $y\in Q\setminus F$ we have $x\not\simul y$ by definition of simulation.

Consider the main iteration of the \textbf{while}-loop and assume that the invariant holds
at the start of the \textbf{while}-loop. We show that it holds after the
\textbf{for}-loops. 

Fix $i$ such that $\NotRel(i)\neq\emptyset$ and $t\in Q$ such that
$\moves{t}{\NotRel(i)}$. Let $\psi=\Reach{t}{\Rel(i)}$, let
$a\in\denote{\neg\psi}$, and let $s\in Q$ be such that $\move(s,a,i)$.
Suppose, by way of contradiction, that $s\simul t$. Then there exists
$j$ such that $\move(t,a,j)$ and $i\simul j$. But, by the definition of
$\psi$ and choice of $a$, it follows that $j\notin\Rel(i)$,
i.e.,\ $(i,j)\in \compl{\Rel}$, and so $i \not\simul j$ follows from
the invariant, which gives us the contradiction. Hence, $s\not\simul
t$ and $\Rel$ is updated by removing $(s,t)$, i.e., $\compl{\Rel}$ as
well as $\NotRel$ gets the new element $(s,t)$.  Thus, the invariant
is preserved, and it follows that ${\simul} \subseteq \Rel$ and, in
particular, that ${\simul_M} \subseteq \Rel$.

We need to show that upon termination $\Rel\subseteq{\simul_M}$.
We define the
\emph{nonsimulation relation of $k$ steps} $\notsim{k}$ over $Q$
by induction over $k$ as follows. We say that \emph{$s$ is
  $k$-nonsimulable by $t$} when $s\notsim{k}t$.
\[
\begin{array}{rll}
  \notsim{0} &\eqdef& F\times (Q\setminus F)\\
  s\notsim{k+1}t &\eqdef& s\notsim{k}t \vee \exists a\exists i( \move(s,a,i)\wedge \forall j (\move(t,a,j)
  \Rightarrow i \notsim{k} j)))
\end{array}
\]
  Then ${\notsim{M}}\eqdef{\notsim{\kappa}}$ where $\kappa$ is such that ${\notsim{\kappa+1}} = {\notsim{\kappa}}$,
  and ${\simul_M} \eqdef {\compl{\notsim{M}}}$.
Thus $s\notsim{M}t$ means that $s$ is $k$-nonsimulable by $t$ for some
$k\geq 0$, or in other words, $t$ cannot $k$-step simulate $s$ for any
$k\geq 0$.
It follows that $\simul_M$ is the unique \emph{maximal simulation} relation of
$M$.
This follows by showing that $\simul_M$ is indeed a simulation
relation, and (by induction over $k$) that for any simulation relation
$\simul$, we have ${\notsim{k}} \subseteq {\compl{\simul}}$.  Hence
${\simul} \subseteq {\simul_M}$.
  
We
show that $\compl{\simul_M} \subseteq \compl{\Rel}$ by showing that
$\notsim{k}\subseteq \compl{\Rel}$ by induction over $k\geq 0$.  This
holds for $k=0$ due to the initial value of $\NotRel$ and the update
of $\Rel$.  Assume now that $(s,t)\in
{\notsim{k+1}\setminus\notsim{k}}$.  Then there exists $a\in\domain$
and $i\in Q$ such that
$\denote{\Delta}(t,a)\subseteq{{\notsim{k}}(i)}$.
Consider the first iteration of the \textbf{while}-loop in which, for
some such $a$ and $i$ with $\move(s,a,i)\in\denote\Delta$, the last
element of ${\notsim{k}}(i)\cap\postofin a t {\denote\Delta}$ is added
to $\compl{\Rel}(i)$ on line \ref{algNoCountSFA:Rel}.
This must indeed eventually happen because
(1) since the automaton is complete, it must hold $\denote\Delta(t,a)$ is not empty,
and (2) due to the induction hypothesis, all elements of ${\notsim{k}}(i)$ eventually appear in $\NotRel(i)$ to be later removed from $\Rel(i)$ on line \ref{algNoCountSFA:Rel}.
Because the last element of ${\notsim{k}}(i)\cap\postofin  a t {\denote\Delta}$ is being removed from $\Rel(i)$, then it is in $\NotRel(i)$, and hence $t \rightarrow \NotRel(i)$.
The transition $t$ is therefore added to $\mathit{Rm}$.  
When $t$ is processed within the \textbf{for}-loop on line~\ref{algNoCountSFA:t},
the satisfiability check of $\neg\psi\wedge\varphi_{si}$ will eventually succeed 
because,
$\denote{\neg\Reach{t}{\Rel(i)}}=\{a\mid
{\denote\Delta}(t,a)\subseteq {\compl{\Rel}}(i)\}$.
And since (by IH) ${\notsim{k}}\subseteq\compl{\Rel}$
it follows that $\denote{\Delta}(t,a)\subseteq\compl{\Rel}(i)$ and so $a\in\denote{\neg\psi}$.
Then $(s,t)$ is added
to $\NotRel$ and deleted from $\Rel$ since all entries that
are added to $\NotRel$ are later deleted from $\Rel$. Since $(s,t)$ was chosen
freely, it follows that $\notsim{k+1}\subseteq \compl{\Rel}$ (by termination of \nocountsim).

%\lukas{ I report that I am convinced by the Correctness proof.  I had a hard
%time to understand the last paragraph, so I tried to change places where I got
%stuck, but feel free to rever my changes.  }

\subsection{Complexity}

Finally, we now proceed to a proof of Lemma~\ref{lemma:nocountsim}.

\begin{proof}[Lemma~\ref{lemma:nocountsim}] The initialization is obviously done
in time $\bigO{n^2}$. 

Further, the construction of the set $\Rm$ is, for a fixed $i$, done on the
whole in $m$ steps. Indeed, we enumerate all transitions going to some $j \in
\NotRel(i)$, and when we sum over all $j \in Q$, we get the number of all
transitions in $M$.  Hence, for all $i \in Q$, the computation is done in
$\bigO{nm}$ time.

For fixed states $i,t \in Q$, the state $t$ can occur in the set $\Rm$ at most
$m_t$ times, and constructing $\psi$ on line~\ref{algNoCountSFA:psi} consists of
iterating through all transitions outgoing from $t$. Therefore, this line is
executed in $\bigO{m_t^2}$ time. Summed over all $i,t \in Q$, we get
$\bigO{n\sum_{t \in Q}m_t^2}$. Since we assume that the time and space
complexity of logical operations are the same, we can assume that the
disjunction involved in the computation of $\Gamma$ has constant complexity and
take it into account later, during the check on
line~\ref{algAbsSFA:check}.\footnote{To be sure that line~\ref{algAbsSFA:check}
is reached at least once, before constructing $\psi$, we can check whether there
are any transitions going into $i$, and if there are not, we continue with the
next iteration of the \textbf{for} loop. For readability of the algorithm, we
have not included this detail into it.} The last fact to show is that
lines~\ref{algAbsSFA:check}--\ref{algNoCountSFA:NotRel:update} are reached at
most $m^2$ times. Let us again fix states $i,t \in Q$. Then,
lines~\ref{algAbsSFA:check}--\ref{algNoCountSFA:NotRel:update} can be reached at
most $m_tm_i^{-1}$ times where $m_i^{-1}$ is the number of transitions going to
$i$. Summing over all states $i,t$, we get $m^2$. 

The predicate $\psi$ is a conjunction of $m_t$ predicates, $n^2 \leq nm \leq
n\sum_{q \in Q}m_q^2$, and so Algorithm \nocountsim has the time complexity
\[\bigO{n\sum_{q \in Q}m_q^2 + m^2\Csat(\maxoutdeg,k)}.\]\qed\vspace{-3mm}\end{proof}

 %%%%%%%%%%%%%%%%%%%%%%%%%%%%%%%%%%%%%%%%%%%%%%%%%%%%%%%%%%%%%%%%%%%%%%%%%%%%%%%%
 \vspace{-0.0mm}
 \section{Simulation vs. Bisimulation-Based Reduction}\label{app:exp}
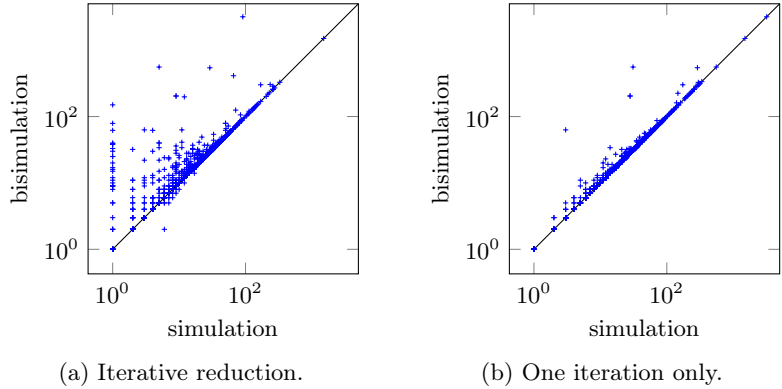
\begin{figure}[t]
 \centering
 \begin{subfigure}{0.45\textwidth}
 \centering
 \begin{tikzpicture}
 \begin{axis}[
     xlabel=simulation,
     ylabel=bisimulation,
     axis equal image,
     ymin=0,
     xmin=0,
     ymax=5000,
     xmax=5000,
     ymode=log,
     xmode=log,
     width=\graphwidthother,
 ]
 \addplot+[mark size=1, only marks,mark=\ourmark] table [x=statesSim, y=statesBisim, col sep=comma] {data/onlythisisused/simvsbisim/Regex.csv};
 \addplot+[mark=none,color=black] plot coordinates {(1,1) (5000,5000)};
 \end{axis}
 \end{tikzpicture}
 \caption{Iterative reduction.}
 \label{app:fig:simbisim-states-iteration}
 \end{subfigure}
 \begin{subfigure}{0.45\textwidth}
 \centering
 \begin{tikzpicture}
 \begin{axis}[
     xlabel=simulation,
     ylabel=bisimulation,
     axis equal image,
     ymin=0,
     xmin=0,
     ymax=5000,
     xmax=5000,
     ymode=log,
     xmode=log,
     width=\graphwidthother,
 ]
 \addplot+[mark size=1, only marks,mark=\ourmark] table [x=statesSim, y=statesBisim, col sep=comma] {data/onlythisisused/simvsbisim/onlyoneRegex.csv};
 \addplot+[mark=none,color=black] plot coordinates {(1,1) (5000,5000)};
 \end{axis}
 \end{tikzpicture}
 \caption{One iteration only.}
 \label{app:fig:simbisim-states-once}
 \end{subfigure}
\caption{Simulation vs. bisimulation-based reduction: the number of states of the reduced automaton.}
\label{app:fig:simbisim-states}
 \end{figure}

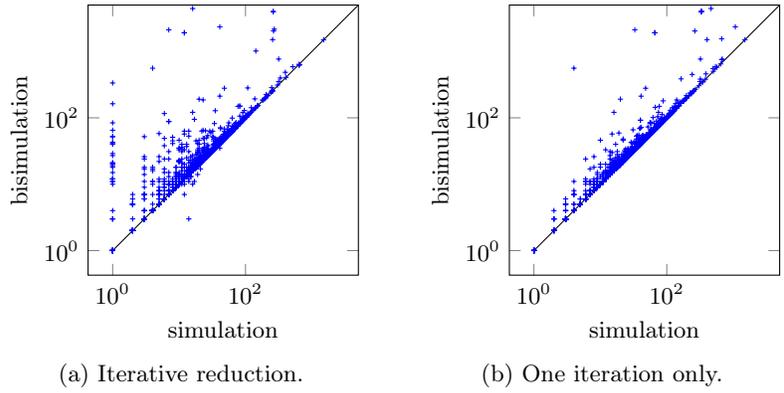
\begin{figure}
 \centering
 \begin{subfigure}{0.45\textwidth}
 \centering
 \begin{tikzpicture}
 \begin{axis}[
     xlabel=simulation,
     ylabel=bisimulation,
     axis equal image,
     ymin=0,
     xmin=0,
     ymax=5000,
     xmax=5000,
     ymode=log,
     xmode=log,
     width=\graphwidthother,
 ]
 \addplot+[mark size=1, only marks,mark=\ourmark] table [x=transitionsSim, y=transitionsBisim, col sep=comma] {data/onlythisisused/simvsbisim/Regex.csv};
 \addplot+[mark=none,color=black] plot coordinates {(1,1) (5000,5000)};
 \end{axis}
 \end{tikzpicture}
 \caption{Iterative reduction.}
 \label{app:fig:simbisim-transitions-iteration}
 \end{subfigure}
 \begin{subfigure}{0.45\textwidth}
 \centering
 \begin{tikzpicture}
 \begin{axis}[
     xlabel=simulation,
     ylabel=bisimulation,
     axis equal image,
     ymin=0,
     xmin=0,
     ymax=5000,
     xmax=5000,
     ymode=log,
     xmode=log,
     width=\graphwidthother,
 ]
 \addplot+[mark size=1, only marks,mark=\ourmark] table [x=transitionsSim, y=transitionsBisim, col sep=comma] {data/onlythisisused/simvsbisim/onlyoneRegex.csv};
 \addplot+[mark=none,color=black] plot coordinates {(1,1) (5000,5000)};
 \end{axis}
 \end{tikzpicture}
 \caption{One iteration only.}
 \label{app:fig:simbisim-transitions-once}
 \end{subfigure}
 \caption{Simulation vs. bisimulation-based reduction: the number of transitions of the reduced automaton.}
\label{app:fig:simbisim-transitions}
\end{figure}

 \vspace{-0.0mm}
 %%%%%%%%%%%%%%%%%%%%%%%%%%%%%%%%%%%%%%%%%%%%%%%%%%%%%%%%%%%%%%%%%%%%%%%%%%%%%%%%
We give a more detailed report on the experimental comparison of the effect of simulation and bisimulation based reduction on the \regex benchmark discussed in Section~\ref{sec:experiments} and also a comparison of the cost of these reductions.  
Fig.~\ref{app:fig:simbisim-states} shows a comparison of the numbers of states of the reduced automata.
The iterative reducing process described in Section~\ref{sec:experiments} is used on Fig.~\ref{app:fig:simbisim-states-iteration}, Fig.~\ref{app:fig:simbisim-states-once} shows the reduction after the first iteration (it corresponds to the ``ordinary'' simulation and bisimulation-based reduction). Fig.~\ref{app:fig:simbisim-transitions} then compares the numbers of transitions. 
One may see that simulation is clearly more powerful and that it may greatly benefit from iterating the forward and backward reduction. 
The comparison of the numbers of states gives a very similar picture as the comparison of the numbers of transitions, but one may see that simulation wins by a slightly larger margin when comparing the numbers of transitions. This is probably due to the use of the removal of simulation smaller transitions, which does not have a meaningful counterpart when working with bisimulations.
%The picture stays essentially the same no matter whether comparing the numbers of states or the numbers of transitions.  

\begin{figure}
 \centering
 \begin{subfigure}{0.45\textwidth}
 \centering
 \begin{tikzpicture}
 \begin{axis}[
     scaled y ticks={base 10:-3},
     ytick scale label code/.code={},
     scaled x ticks={base 10:-3},
     xtick scale label code/.code={},
     xlabel=simulation,
     ylabel=bisimulation,
     axis equal image,
     ymin=0,
     xmin=0,
     ymode=log,
     xmode=log,
     width=\graphwidthother,
 ]
 \addplot+[mark size=0.3, only marks,mark=\ourmark] table [x=timeSim, y=timeBisim, col sep=comma] {data/onlythisisused/simvsbisim/Regex.csv};
 \addplot+[mark=none,color=black] plot coordinates {(0.01,0.01) (40000,40000)};
 \end{axis}
 \end{tikzpicture}
 \caption{Iterative reduction.}
 \label{app:fig:simbisim-time-iteration}
 \end{subfigure}
 \begin{subfigure}{0.45\textwidth}
 \centering
 \begin{tikzpicture}
 \begin{axis}[
     scaled y ticks={base 10:-3},
     ytick scale label code/.code={},
     scaled x ticks={base 10:-3},
     xtick scale label code/.code={},
     xlabel=simulation,
     ylabel=bisimulation,
     axis equal image,
     ymin=0,
     xmin=0,
     ymode=log,
     xmode=log,
     width=\graphwidthother,
 ]
 \addplot+[mark size=0.3, only marks,mark=\ourmark] table [x=timeSim, y=timeBisim, col sep=comma] {data/onlythisisused/simvsbisim/onlyoneRegex.csv};
 \addplot+[mark=none,color=black] plot coordinates {(0.01,0.01) (40000,40000)};
 \end{axis}
 \end{tikzpicture}
 \caption{One iteration only.}
 \label{app:fig:simbisim-time-once}
 \end{subfigure}
 \caption{Simulation vs. bisimulation-based reduction: runtime in miliseconds. }
 \label{app:fig:simbisim-time}
 \end{figure}
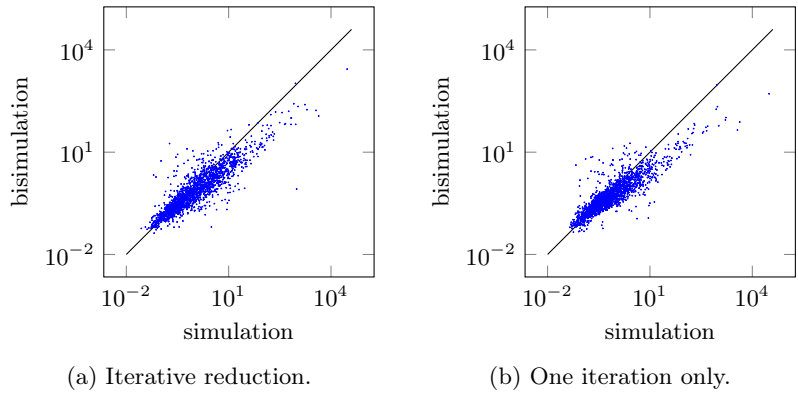

Lastly, Fig.~\ref{app:fig:simbisim-time} shows a comparison of the running times of the simulation and bisimulation reduction. 
Fig.~\ref{app:fig:simbisim-time-iteration} shows the overall time needed by the iterative reduction process, Fig.~\ref{app:fig:simbisim-time-once} then the time taken by the first iteration---essentially the time taken by computing the simulation preorder or the bisimulation equivalence. 
One may see that bisimulation is cheaper overall, especially when the automata are growing larger (note the logarithmic scale). 
However, computing simulation may be faster in surprisingly many cases.

%%%%%%%%%%%%%%%%%%%%%%%%%%%%%%%%%%%%%%%%%%%%%%%%%%%%%%%%%%%%%%%%%%%%%%%%%%%%%%%%
\end{document}